\documentclass[a4paper,11pt]{article}

\usepackage{jheppub} 
\usepackage{graphicx,color}
\usepackage{amsmath}
\usepackage{slashed}
\usepackage{multirow}
\usepackage{autobreak}
\usepackage[normalem]{ulem}
\usepackage{hyperref}
\usepackage{cleveref}
\usepackage{slashed}
\usepackage{amssymb}
\usepackage{epsfig}
\usepackage{array}
\usepackage{mathtools}
\usepackage{caption}
\usepackage[utf8]{inputenc}
\usepackage{mathtools, nccmath}
\usepackage{float}
\usepackage[section]{placeins}
\usepackage[utf8]{inputenc}
    \usepackage[T1]{fontenc}
    \usepackage{lmodern}
    \usepackage[x11names]{xcolor}
    \usepackage{framed}
\usepackage{tikz-feynman}
\colorlet{shadecolor}{blue!10}

\tikzfeynmanset{every dot/.style={minimum size=5pt}}

\SetSymbolFont{largesymbols}{bold}{OMX}{txex}{b}{n}

\DeclareMathAlphabet{\mathpzc}{OT1}{pzc}{m}{it}

\newcommand{\be}{\begin{eqnarray*}}
\newcommand{\ee}{\end{eqnarray*}}

\usepackage{parskip}

\newcommand{\ba}{\begin{array}}
\newcommand{\ea}{\end{array}}
\newcommand{\bd}{\begin{displaymath}}
\newcommand{\ed}{\end{displaymath}}
\newcommand{\beq}{\begin{equation}}
\newcommand{\eeq}{\end{equation}}

\def\vevof#1{\left\langle #1  \right\rangle}

\def\q2 {q^2}

\def\bt{\begin{table}}
\def\et{\end{table}}

	\title{\boldmath Collider and CMB complementarity of leptophilic dark matter with light Dirac neutrinos}

\author{Debasish Borah,}
\author{Nayan Das,}
\author{Sahabub Jahedi,}
\author{Bhavya Thacker}

\affiliation{Department of Physics, Indian Institute of Technology Guwahati, Assam 781039, India}

\emailAdd{dborah@iitg.ac.in}
\emailAdd{nayan.das@iitg.ac.in}
\emailAdd{sahabubjahedi@gmail.com}
\emailAdd{b.thacker172@gmail.com}

\abstract{We study the discovery prospects of leptophilic dark matter (DM) in future lepton colliders by considering the light neutrinos to be of Dirac type. Adopting an effective field theory (EFT) approach, we write down dimension six operators connecting the standard model (SM) fields, light Dirac neutrinos and DM. Considering DM relic to be generated via the thermal freeze-out, we check the discovery prospects at future lepton colliders via mono-photon plus missing energy searches. The right chiral parts of light Dirac neutrinos get thermalised due to their interactions with the bath as well as leptophilic DM, leading to enhanced effective relativistic degrees of freedom $N_{\rm eff}$ within reach of future cosmic microwave background (CMB) experiments. The interplay of existing bounds from cosmological observations related to DM relic and $N_{\rm eff}$, direct and indirect detection of DM, astrophysics, and collider observations leave promising discovery prospects at future electron and muon colliders along with complementary signatures at future CMB experiments.}


\begin{document}

\maketitle
\flushbottom

\section{Introduction}
\label{sec:intro}
The presence of dark matter (DM) in our Universe has been suggested by several irrefutable evidences in astrophysics and cosmology related observations \cite{Zyla:2020zbs, Planck:2018vyg}. However, the particle nature of DM and its non-gravitational interactions are not yet known. While none of the Standard Model (SM) particles can be a DM candidate, several beyond Standard Model (BSM) proposals exist in the literature among which the weakly interacting massive particle (WIMP) remains the most widely studied one. In the WIMP paradigm, a particle DM candidate having mass and interaction strength typically around the electroweak ballpark can give rise to the observed DM abundance after thermal freeze-out, a remarkable coincidence often referred to as the \textit{WIMP Miracle}. Similar to DM, the observed neutrino mass and mixing also indicate the presence of BSM physics. Assuming the neutrinos to acquire masses solely from the SM Higgs, one needs to introduce the right chiral neutrinos, singlet under the SM gauge symmetry. If we forbid Majorana mass of right-handed neutrinos, pure Dirac neutrino mass can arise from SM Higgs alone, though at the cost of fine-tuned Yukawa couplings. Several UV complete Dirac neutrino models exist in the literature where such fine-tuning can be avoided at the cost of enlarged symmetries and particle content \cite{Roncadelli:1983ty, Roy:1983be, Babu:1988yq,
	Peltoniemi:1992ss, Chulia:2016ngi, Aranda:2013gga, 
	Chen:2015jta, Ma:2015mjd, Reig:2016ewy, Wang:2016lve, Wang:2017mcy, Wang:2006jy, 
	Gabriel:2006ns, Davidson:2009ha, Davidson:2010sf, Bonilla:2016zef, 
	Farzan:2012sa, Bonilla:2016diq, Ma:2016mwh, Ma:2017kgb, Borah:2016lrl, 
	Borah:2016zbd, Borah:2016hqn, Borah:2017leo, CentellesChulia:2017koy, 
	Bonilla:2017ekt, Memenga:2013vc, Borah:2017dmk, CentellesChulia:2018gwr, 
	CentellesChulia:2018bkz, Han:2018zcn, Borah:2018gjk, Borah:2018nvu, Borah:2019bdi, CentellesChulia:2019xky, Jana:2019mgj, Nanda:2019nqy, Guo:2020qin, Bernal:2021ezl, Borah:2022obi, Li:2022chc, Dey:2024ctx, Singh:2024imk, Borah:2024gql}.

In this work, we consider DM to be of WIMP-type with light neutrinos being Dirac fermions. While DM can be of any spin, we consider it to be a Dirac fermion $(\chi)$, for simplicity. We adopt an effective field theory (EFT) approach where all BSM interactions are written in terms of higher dimensional effective operators. The interactions of DM, as well as the right chiral part of Dirac neutrino $(\nu_R)$ with the SM and among themselves, are suppressed by some powers of the same cutoff scale $\Lambda$. The phenomenology in such an EFT setup can be discussed with very few free parameters, namely, DM mass, cutoff scale, and the corresponding coupling or Wilson coefficients. DMEFT has been studied extensively in the context of direct detection, indirect detection as well as collider searches in several works \cite{Beltran:2008xg, Fan:2010gt, Goodman:2010ku, Beltran:2010ww, Fitzpatrick:2012ix}, also summarised in a recent review \cite{Bhattacharya:2021edh}. EFT of light Dirac neutrino interactions has also been studied in the context of $N_{\rm eff}$ \cite{Luo:2020sho, Luo:2020fdt}. EFT of light species, in general, contributing to $N_{\rm eff}$ was discussed in \cite{Brust:2013ova}. Similarly, EFT of non-standard neutrino interactions without DM or other light species was discussed in the context of $N_{\rm eff}$ by the authors of \cite{Du:2021idh}. For the first time, we consider DM and light Dirac neutrinos in an EFT framework to study the complementary detection prospects at collider and CMB experiments. We also consider the DM to be leptophilic such that DM interacts only with SM leptons and $\nu_R$. While this helps in evading strong direct-detection bounds \cite{LZ:2022lsv,LZ:2024} to some extent, it also opens up interesting discovery prospects of DM at lepton colliders like the $e^+ e^-$ colliders \cite{Fox:2011fx, Essig:2013vha, Yu:2013aca, Kadota:2014mea, Yu:2014ula, Freitas:2014jla, Dutta:2017ljq, Liu:2019ogn, Choudhury:2019sxt, Kundu:2021cmo, Barman:2021hhg, Bhattacharya:2022wtr, Bhattacharya:2022qck, Ge:2023wye, Roy:2024ear, Barman:2024nhr} as well as muon colliders \cite{Han:2020uak}. Considering light neutrinos to be of Dirac type, we have additional DM annihilations, DM direct/indirect detection channels, and collider backgrounds. We include these to constrain the parameter space using bounds from cosmological observation of DM relic, searches in collider, direct, indirect detection experiments as well as indirect astrophysical bounds from compact objects. We then estimate the contribution of thermalised $\nu_R$ to $N_{\rm eff}$ to put additional constraints on the parameter space from cosmic microwave background (CMB) constraints ${\rm N_{eff}= 2.99^{+0.34}_{-0.33}}$ at $2\sigma$ or $95\%$ CL including baryon acoustic oscillation (BAO) data \cite{Planck:2018vyg}. Depending upon the specific $\nu_R$-SM or $\nu_R$-DM operators, the CMB bound can be stronger or weaker than terrestrial bounds. While future lepton colliders can probe some part of the available parameter space allowed by current data, future CMB experiments like CMB Stage IV (CMB-S4) \cite{Abazajian:2019eic}, CMB-HD \cite{CMB-HD:2022bsz} can probe the entire parameter space consistent with thermal DM. 

This paper is organised as follows. In section \ref{sec1}, we list out the effective operators to be used in our study of DM and $N_{\rm eff}$. In section \ref{sec2}, we discuss DM phenomenology, including direct and indirect signatures, followed by the details of collider phenomenology in section \ref{sec3}. In section \ref{sec4}, we discuss the details of $N_{\rm eff}$ followed by brief remarks on UV completions of the effective operators in section \ref{sec4a}. We finally conclude in section \ref{sec5}.

\section{Effective Operators}
\label{sec1}
We consider the DM to be a Dirac fermion $\chi$ singlet under the SM gauge symmetry. An additional $Z_2$ symmetry under which $\chi$ is odd while all other fields are even, stabilises DM. We also consider three copies of right chiral neutrino $\nu_R$, which constitute Dirac fermion with the left-chiral counterpart in the SM. We also consider DM to be leptophilic and then write down the EFT operators of the form $\mathcal{O}_{\rm DM} \mathcal{O}_{\rm SM}$, $\mathcal{O}_{\rm DM} \mathcal{O}_{\nu_R}$, and $\mathcal{O}_{\rm SM-\nu_R}$, suppressed by appropriate powers of the cutoff scale $\Lambda$. All these operators are invariant under SM gauge symmetry. In addition to $Z_2$ symmetry stablising DM, we also consider a global lepton number symmetry which prevents Majorana mass term of neutrinos guaranteeing its Dirac nature. We also do not include baryon number violating operators in order to avoid strong bounds from nucleon lifetime.

With these considerations, the operators of the lowest dimensions, namely of dimension 6 connecting DM-SM and DM-$\nu_R$, are listed in Table~\ref{tab:dm.ops}. Similarly, dimension-6 operators $\mathcal{O}_{\rm SM-\nu_R}$connecting SM with $\nu_R$ are presented in Table~\ref{tab:rhn.ops}. Here $L, Q$ denote lepton, quark doublets whereas $e_R, u_R, d_R$ denote corresponding $SU(2)$ singlet charged fermions of the SM. $H$ denotes the Higgs doublet while $W^{\mu \nu}, B^{\mu \nu}$ denote $SU(2), U(1)_Y$ field strength tensors, respectively. It should be noted that all these operators lead to interaction Lagrangian terms of the type 
\begin{equation}
	\frac{c_{\chi l}}{\Lambda^2} \mathcal{O}_{\chi l}, \,\, \frac{c_{\chi \nu}}{\Lambda^2} \mathcal{O}_{\chi \nu},  \,\, \frac{c_{\rm SM \nu_R}}{\Lambda^2} \mathcal{O}_{\rm SM-\nu_R}
\end{equation}
with $c_{ij}$ being the corresponding Wilson coefficients. While we consider all leptophilic DM-SM operators of dimension 6 in our analysis, the $\nu_R$-SM and $\nu_R$-DM operators are considered case by case to check the CMB bounds. As we will discuss in details in the upcoming sections, for certain classes of $\nu_R$-SM operators, current CMB bounds can overshadow all existing constraints as well as future sensitivities while for some other types of $\nu_R$-SM or $\nu_R$-DM operators such bounds can be relatively weaker. This gives rise to an interesting complementarity among future collider and CMB bounds on leptophilic DM parameter space when light neutrinos are of Dirac nature. Light Dirac neutrino mass arises from renormalisable coupling of lepton doublet $L$, $\nu_R$ with the SM Higgs doublet $H$. Since the corresponding Dirac Yukawa couplings are very small $\lesssim \mathcal{O}(10^{-12})$ in order to be consistent with sub-eV neutrino mass, they do not play any significant role in collider as well as $N_{\rm eff}$ phenomenology. Therefore, the flavour structure of Dirac Yukawa coupling can be chosen to fit neutrino oscillation data without affecting our analysis here. 

\begin{table}[h]
	\centering
	\begin{tabular}{|c|c|c|c|}
		\hline
		\textbf{Class} & \textbf{Name} & $\mathcal{O}_{\rm DM}$ & $\mathcal{O}_{\rm SM}, \mathcal{O}_{\nu_R}$\\
		\hline
		DMEFT & $\mathcal{O}_{\chi \ell}$ & $(\overline{\chi} \gamma^\mu \chi), (\overline{\chi} \gamma^\mu \gamma^5 \chi)$ & $(\overline{L} \gamma_\mu L), (\overline{e_R} \gamma_\mu e_R)$ \\
		\hline
		$\nu$DMEFT & $\mathcal{O}_{\chi \nu}$ & $(\overline{\chi} \gamma^\mu \chi), (\overline{\chi} \gamma^\mu \gamma^5 \chi)$ & $(\overline{\nu_R} \gamma_\mu \nu_R)$ \\
		\hline
	\end{tabular}
	\caption{Operators involving DM, SM, and $\nu_R$ leading to interactions at dimension 6 level relevant for leptophilic DM phenomenology.}
	\label{tab:dm.ops}
\end{table}

\begin{table}[h]
	\centering
	\begin{tabular}{|c|c|c|c|c|c|}
		\hline
		\textbf{No.} & \textbf{Name} & \textbf{Operator} & \textbf{No.} & \textbf{Name} & \textbf{Operator} \\
		\hline
		1 & $\mathcal{O}_{L \nu H}$ & $(\overline{L} \nu_R) \tilde{H} (H^\dagger H) + \text{h.c.}$ & 9 & $\mathcal{O}_{d \nu}$ & $(\overline{d_R} \gamma^\mu d_R) (\overline{\nu_R} \gamma_\mu \nu_R)$ \\
		\hline
		2 & $\mathcal{O}_{H \nu}$ & $(\overline{\nu_R} \gamma^\mu \nu_R) (H^\dagger i \overleftrightarrow{D}_\mu H)$ & 10 & $\mathcal{O}_{d u \nu e}$ & $(\overline{d_R} \gamma^\mu u_R) (\overline{\nu_R} \gamma_\mu e_R) + \text{h.c.}$ \\
		\hline
		3 & $\mathcal{O}_{H \nu e}$ & $(\overline{\nu_R} \gamma^\mu e_R) (\tilde{H}^\dagger i D_\mu H) + \text{h.c.}$ & 11 & $\mathcal{O}_{L \nu}$ & $(\overline{L} \gamma^\mu L) (\overline{\nu_R} \gamma_\mu \nu_R)$ \\
		\hline
		4 & $\mathcal{O}_{\nu B}$ & $(\overline{L} \sigma_{\mu\nu} \nu_R) \tilde{H} B^{\mu\nu}$ & 12 & $\mathcal{O}_{Q \nu}$ & $(\overline{Q} \gamma^\mu Q) (\overline{\nu_R} \gamma_\mu \nu_R)$ \\
		\hline
		5 & $\mathcal{O}_{\nu W}$ & $(\overline{L} \sigma_{\mu\nu} \nu_R) \tau^I \tilde{H} W^{I \mu\nu}$ & 13 & $\mathcal{O}_{L \nu Le}$ & $(\overline{L}^i \nu_R) \epsilon_{ij} (\overline{L}^j e_R) + \text{h.c.}$ \\
		\hline
		6 & $\mathcal{O}_{\nu \nu}$ & $(\overline{\nu_R} \gamma^\mu \nu_R) (\overline{\nu_R} \gamma_\mu \nu_R)$ & 14 & $\mathcal{O}_{L \nu Qd}$ & $(\overline{L}^i \nu_R) \epsilon_{ij} (\overline{Q}^j d_R) +\text{h.c.}$ \\
		\hline
		7 & $\mathcal{O}_{e \nu}$ & $(\overline{e_R} \gamma^\mu e_R) (\overline{\nu_R} \gamma_\mu \nu_R)$ & 15 & $\mathcal{O}_{LdQ \nu}$ & $(\overline{L}^i d_R) \epsilon_{ij} (\overline{Q}^j \nu_R) +\text{h.c.}$ \\
		\hline
		8 & $\mathcal{O}_{u \nu}$ & $(\overline{u_R} \gamma^\mu u_R) (\overline{\nu_R} \gamma_\mu \nu_R)$ & 16 & $\mathcal{O}_{Qu \nu L}$ & $(\overline{Q}^{i} u_R) (\overline{\nu_R} L^i) + \text{h.c.}$ \\
		\hline
	\end{tabular}
	\caption{Dimension 6 operators for SM-$\nu_R$ interactions considering conservation of lepton and baryon numbers $\Delta L = 0, \Delta B = 0$ \cite{Bhattacharya:2015vja,Liao:2016qyd}.}
	\label{tab:rhn.ops}
\end{table}

\section{Dark Matter Phenomenology}
\label{sec2}
\subsection{Relic density}
We calculate DM relic by considering it to be in thermal equilibrium in the early universe followed by thermal freeze-out at an epoch when DM-SM interactions fall out of equilibrium \cite{Kolb:1990vq}. We consider all Wilson coefficients related to DM interactions shown in table \ref{tab:dm.ops} to be $\mathcal{O}(1)$ such that we are left with only two free parameters to constrain from successful DM phenomenology namely, DM mass $m_\chi$ and the cutoff scale $\Lambda$. The Boltzmann equation for DM assuming $2 \rightarrow 2$ annihilations can be written as
\beq
\frac{d Y_\chi}{d x} = -1.32\,\sqrt{g_*}~M_{\text{pl}}~ \frac{m_{\chi} }{x^2}~\vevof{\sigma v}_{2_{\rm DM}\to 2_{\rm SM}} ~\left( {Y_\chi}^2-Y_{\tt eq}^{2} \right)\,, 
\label{eq:beq}
\eeq
\noindent where $M_{\tt pl}$ represents the reduced Planck mass, $Y_\chi = n/s$ denotes the comoving number density of DM with $n, s=g_{*s} \frac{2\pi^2}{45} T^3$ being DM number density and entropy density of the Universe respectively. The dimensionless variable $x$ is defined as $x= m_{\chi} /T$ with $T$ being the temperature. $Y_{\tt eq}$ indicates the value of $Y_\chi$ in thermal equilibrium, described by the Maxwell-Boltzmann distribution for non-relativistic species
\beq
Y_{\tt eq}(x)=0.145~\frac{g_{\tt DM}}{g_{*s}} x^{3/2}e^{-x}\,.
\label{eq:yeq}
\eeq
In Eq.~\eqref{eq:yeq}, $g_{\tt DM}$ indicates the number of internal degrees of freedom for DM while $g_*$ represents the effective relativistic degrees of freedom defined as \cite{Gondolo:1990dk}
\begin{equation}
	g^{1/2}_* = \frac{g_{*s}}{g^{1/2}_{*\rho}} \left( 1+\frac{T}{3g_{*s}} \frac{dg_{*s}}{dT} \right).
\end{equation}
Here $g_{*\rho}$ denotes the relativistic degrees of freedom in energy density of the Universe $\rho = g_{*\rho} \frac{\pi^2}{30} T^4$. For heavy DM freezing out above the electroweak scale, we have $g_* = g_{*s}=g_{*\rho}$. In Eq. \eqref{eq:beq}, $\vevof{\sigma v}_{2_{\rm DM}\to 2_{\rm SM}}$ denotes the thermally averaged annihilation cross-section for $\chi \overline{\chi} \rightarrow \rm SM \, SM$ \cite{Gondolo:1990dk, Guo:2010hq}.

\begin{figure}[h]
	\centering
	\includegraphics[scale=0.5]{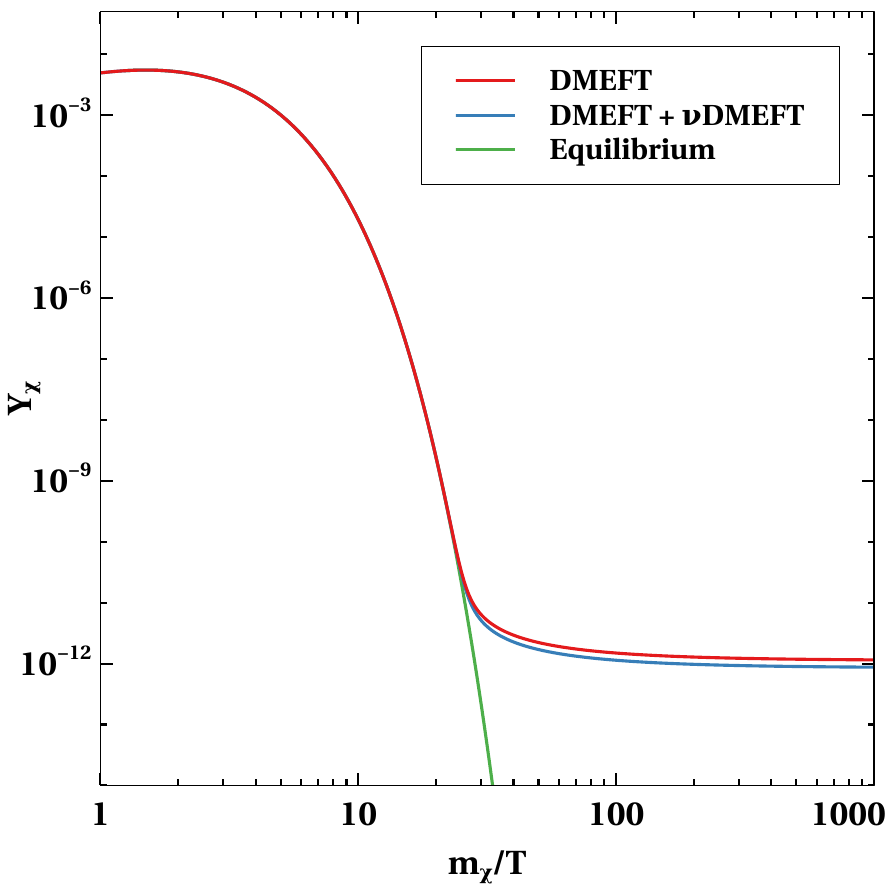}~
	\includegraphics[scale=0.5]{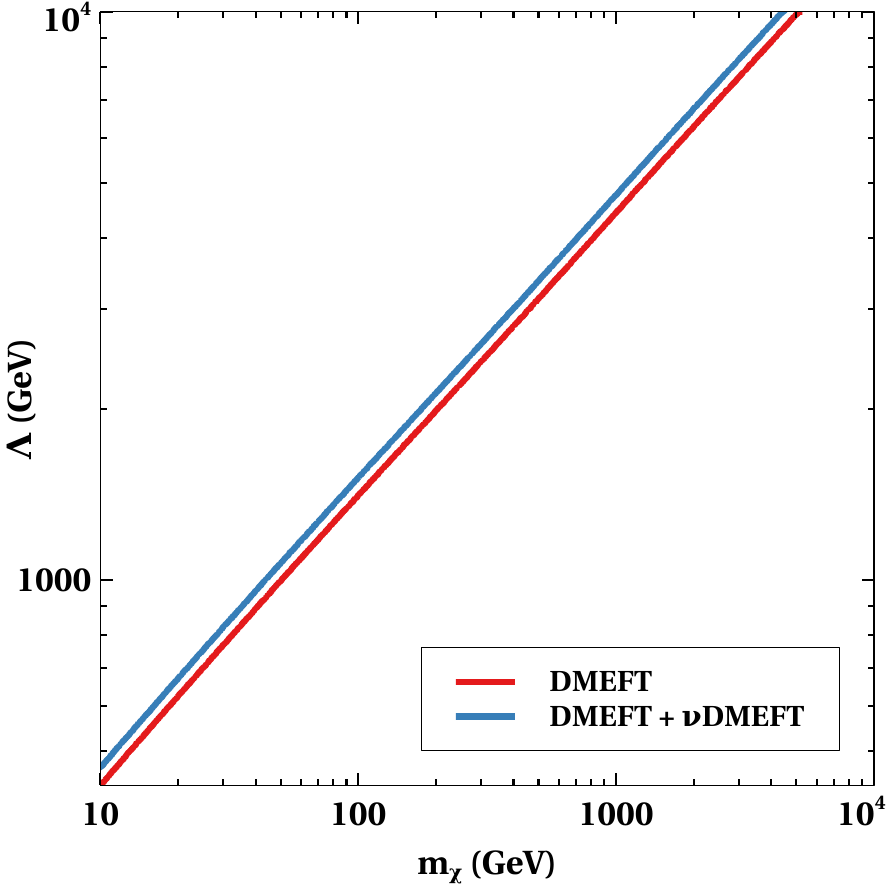}
	\caption{Left panel: DM yield variation with $x=m_{\chi}/T$ with $m_{\chi}=260$ GeV and $\Lambda=2500$ GeV. Right panel: Relic allowed parameter space in $\Lambda-m_{\chi}$ plane.}
	\label{fig:relic}
\end{figure}

In Eq.~\eqref{eq:beq},  $Y_\chi=Y_{\tt eq}$ at $x \to 0$, guarantees DM to be in the thermal bath initially. The relic density of DM is calculated by
\beq
\Omega_{\chi} h^2 = \frac{(1.09 \times 10^9~\text{GeV}^{-1})x_f}{\sqrt{g_*} M_{\text{pl}}} \left( \int^{\inf}_{x_f} dx \frac{\vevof{\sigma v}}{x^2}\right)^{-1},
\label{eq:relic}
\eeq
with $\Omega_{\chi}=\rho_{\text{DM}}/\rho_c$ with $\rho_{\text{DM}}$, $\rho_c$ being the DM energy density and the critical density of the Universe, respectively. The freeze-out epoch is denoted by $x_f=m_{\chi}/T_f$, $T_f$ indicating the freeze-out temperature. Following Eq.~\eqref{eq:beq}, we show the evolution of comoving number density of DM in $Y_{\chi}-x$ plane in the left plot of Fig.~\ref{fig:relic}. Clearly, addition of new annihilation channels from $\mathcal{O}_{\chi \nu}$ operators, DM freeze-out occurs later, leaving a smaller relic. From the right panel plot in Fig.~\ref{fig:relic}, it is evident that as DM mass increases, $\Lambda$ has to grow to maintain the relic abundance as followed from $\Omega_{\chi} h^2 \propto 1/\vevof{\sigma v} \propto \Lambda^4/m_{\chi}^2$. After inclusion of $\mathcal{O}_{\chi \nu}$ operators, the relic allowed parameter space is shifted upward due to the increase in annihilation channels.

\subsection{Direct search}
\begin{figure}[htb!]
	\centering
	\includegraphics[scale=0.5]{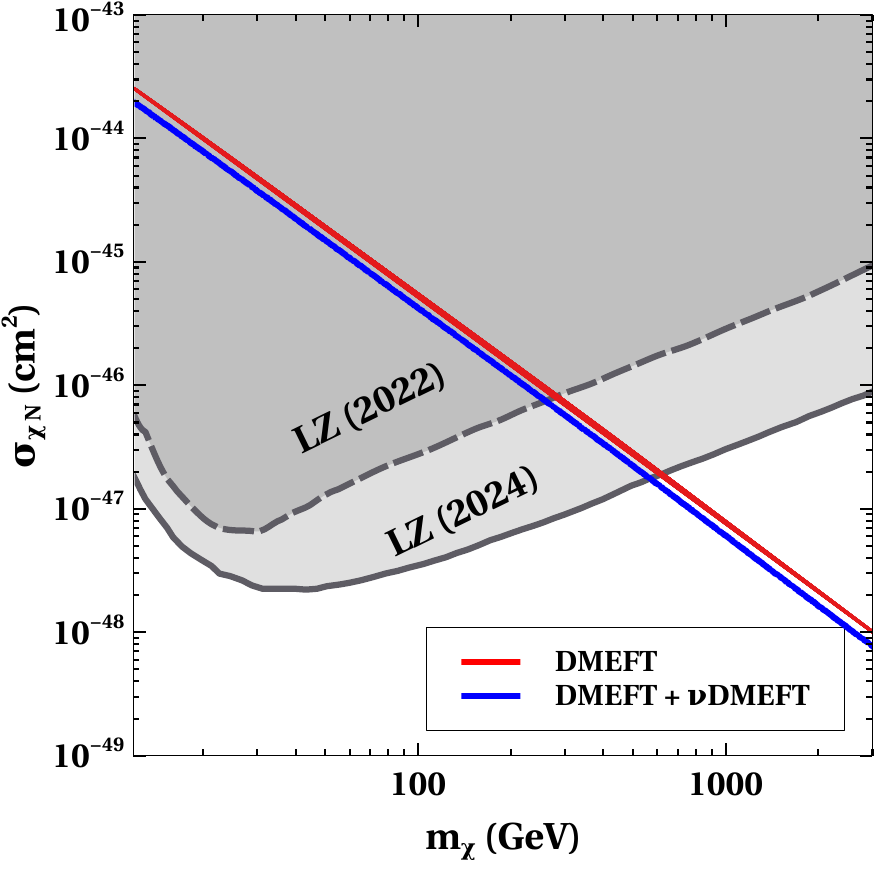}
	\caption{Spin-independent DM-nucleon cross-section for relic density allowed parameter space denoted by red (DMEFT) and blue (DMEFT+$\nu$DMEFT) contours. Dark gray (light gray) shaded region is excluded by LZ experiment bound published in 2022 \cite{LZ:2022lsv} (2024 \cite{LZ:2024}) in $\sigma_{\chi N}-m_{\chi}$ plane.}
	\label{fig:dd}
\end{figure}

Direct detection of DM involves observing nuclear/electron recoil events resulting from DM particles scattering off nucleons or electrons at the earth-based detectors. Though leptophilic DM couples to leptons directly, for the range of DM mass considered in our work, DM-electron scattering is not constrained. On the other hand, such DM does not couple to quarks or nucleons directly but such interactions arise through photon and $Z$ exchange at the one-loop level, as illustrated in the left panel of Fig.~\ref{fig:dd.diag}. In this loop, any charged lepton that the DM interacts with can be present. At direct detection scale $\Lambda_{\text{DD}}$, the dominant spin-independent DM-nucleon cross-section pertaining to $\mathcal{O}_{\chi \ell}$ operators with photon mediation\footnote{The $Z$ mediation has negligible contribution to the DD cross-section due to $Z$ boson mass suppression \cite{Kopp:2009et}.} is expressed as
\beq
\sigma_{\chi N} \approx \frac{\mu_p^2 }{9\pi A^2}\Biggl(\frac{\alpha_\text{em} \mathcal{Z}}{\pi \Lambda^2}\Biggr)^2\Biggl[\ln\Biggl(\frac{\Lambda^2}{\Lambda_\text{DD}^2}\Biggr)\Biggr]^2  \,,
\eeq
\begin{figure}[htb!]
	\centering
	\begin{tikzpicture}[baseline={(current bounding box.center)}, style={scale=0.7, transform shape}]
		\begin{feynman}
			\vertex  (a);
			\vertex [above left=2.8cm of a] (b) {\Large $\chi$};
			\vertex [above right=2.8cm of a] (c) {\Large $\chi$};
			\vertex [below=2.5cm of a] (x) ;
			\vertex [below=2cm of x] (d) ;
			\vertex [below left=2.8cm of d] (e) {\Large $N$};
			\vertex [below right=2.8cm of d] (f) {\Large $N$};
			
			\diagram{
				(b) -- [fermion, ultra thick, arrow size=2.2pt] (a) -- [fermion, ultra thick, arrow size=2.2pt] (c),
				(a) -- [fermion, half left, looseness=1.7, ultra thick, arrow size=2.2pt, edge label=\Large{$\ell$}] (x) -- [fermion, half left, looseness=1.7, ultra thick, arrow size=2.2pt, edge label=\Large{$\bar{\ell}$}] (a),
				(x) -- [boson, ultra thick, arrow size=2.2pt, edge label=\Large{$\gamma, Z$}] (d),
				(d) -- [anti fermion, ultra thick, arrow size=2.2pt] (e),
				(d) -- [fermion, ultra thick, arrow size=2.2pt] (f),
			};
			\draw[fill=black] (a) circle(3mm);
			\draw[fill=white] (d) circle(3mm);
			
		\end{feynman}
	\end{tikzpicture} \quad
	\begin{tikzpicture}[baseline={(current bounding box.center)}, style={scale=0.7, transform shape}]
		\begin{feynman}
			\vertex  (a);
			\vertex [above left=2.8cm of a] (b) {\Large $\chi$};
			\vertex [above right=2.8cm of a] (c) {\Large $\chi$};
			\vertex [below=2.5cm of a] (x) ;
			\vertex [below left=2.8cm of x] (e) {\Large $N$};
			\vertex [below right=2.8cm of x] (f) {\Large $N$};
			
			\diagram{
				(b) -- [fermion, ultra thick, arrow size=2.2pt] (a) -- [fermion, ultra thick, arrow size=2.2pt] (c),
				(a) -- [fermion, half left, looseness=1.7, ultra thick, arrow size=2.2pt, edge label=\Large{$\nu_R$}] (x),
				(x) -- [fermion, half left, looseness=1.7, ultra thick, arrow size=2.2pt, edge label=\Large{$\bar{\nu}_R$}] (a),
				(x) -- [anti fermion, ultra thick, arrow size=2.2pt] (e),
				(x) -- [fermion, ultra thick, arrow size=2.2pt] (f),
			};
			\draw[fill=black] (a) circle(3mm);
			\draw[fill=black] (x) circle(3mm);
		\end{feynman}
	\end{tikzpicture}
	\caption{1-loop Feynman diagrams from DM-nucleon scattering. Left: DMEFT contribution; right: $\nu$DMEFT contribution. }
	\label{fig:dd.diag}
\end{figure}
\noindent
where $\mu_p=m_{\chi} m_p/(m_{\chi}+m_p)$ represents the reduced mass, $\alpha_{\text{em}}$ denotes the fine structure constant, and $A, \mathcal{Z}$ correspond to the mass number and atomic number of the xenon target, respectively. The details of the one-loop integral and amplitude of the DM nucleon cross-section can be found in \cite{Kopp:2009et}. The $\mathcal{O}_{\nu \ell}$ operator also contributes to the dark matter-nucleon cross-section through a one-loop tadpole diagram (right panel diagram of Fig.~\ref{fig:dd.diag}). However, this contribution is suppressed by \(1/\Lambda^8\), making its impact on direct detection negligible. The non-observation of DM signal in detectors imposes the strongest constraints on the DM-nucleon cross-section, with the most stringent limits arising from the LUX-ZEPLIN experiment \cite{LZ:2022lsv}. In Fig.~\ref{fig:dd}, we show relic allowed parameter space in $\sigma_{\chi N}-m_{\chi}$ plane. For the DMEFT scenario, the ruled-out parameter space is up to $(m_{\chi}, \Lambda) \sim (284, 2390) \, \text{GeV}$. After including $\nu$DMEFT, the ruled out limits shift to $(m_{\chi}, \Lambda) \sim (257, 2432) \, \text{GeV}$. Considering the latest spin-independent direct search bound from LZ experiment \cite{LZ:2024}, below $(m_{\chi}, \Lambda) \sim (630, 3500) \, \text{GeV}$ for DMEFT scenario is ruled out. With the inclusion of $\nu$DMEFT, this limit changes to $(m_{\chi}, \Lambda) \sim (570, 3600) \, \text{GeV}$. Clearly, the inclusion of right-handed neutrinos relaxes the direct detection bounds on DM to some extent.

\subsection{Indirect search}
\begin{figure*}[htb!]
	\centering
	\includegraphics[scale=0.5]{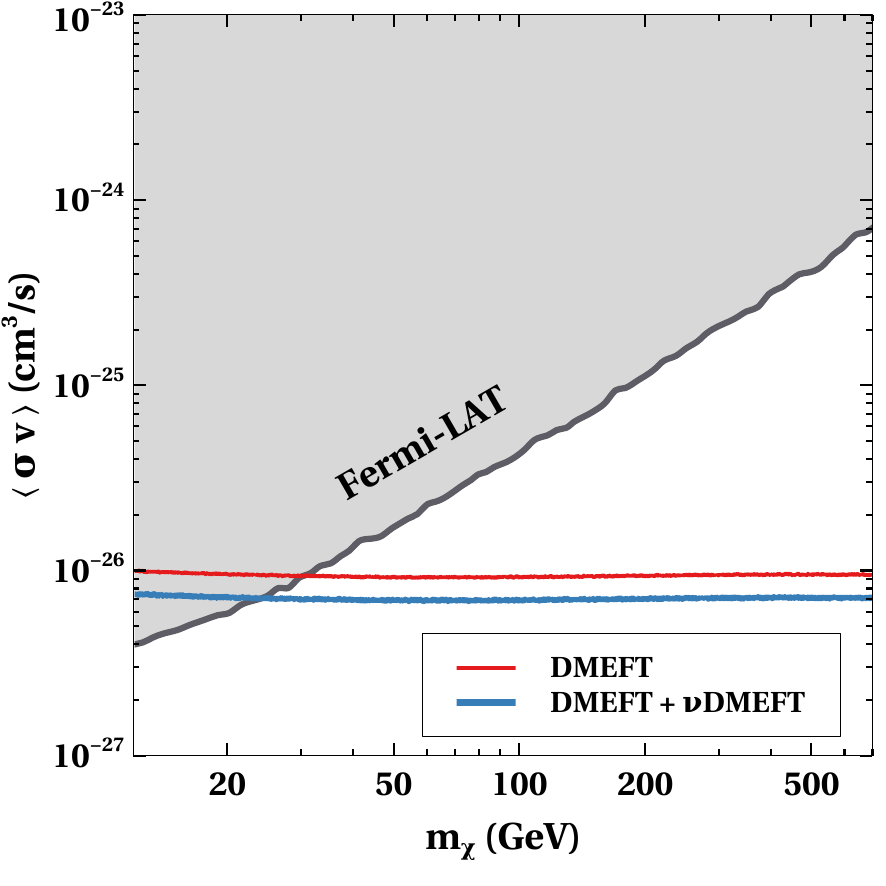}
	\caption{Relic allowed parameter space in $\vevof{\sigma v}-m_{\chi}$ plane denoted by red (DMEFT) and blue (DMEFT+$\nu$DMEFT) contours. Gray shaded region is excluded from combined data analysis of MAGIC Cherenkov telescopes and Fermi-LAT experiment \cite{Fermi-LAT:2016afa} in $\vevof{\sigma v}_{\chi \chi \rightarrow \tau^+ \tau^-}$ versus $m_{\chi}$ plane.}
	\label{fig:id}
\end{figure*}

WIMP-type DM also has promising indirect detection prospects due to its annihilation into SM particles 
$\chi \chi \to f \bar{f}, \gamma \gamma, ...$, where $f$'s are the SM fermions. The excess of antimatter or photons (diffuse and monochromatic) can be constrained from observations made by numerous satellites like the Fermi Large Area Telescope (Fermi-LAT) \cite{Fermi-LAT:2016afa} or future ground-based telescopes like the Cherenkov Telescope Array (CTA) \cite{CTAConsortium:2012fwj}. For stable DM, such observations constrain the annihilation cross-section of DM $\vevof{\sigma v}$ into specific final states. The non-observation of any gamma-ray excess by Fermi-LAT imposes an upper bound on DM annihilation into charged fermions like $\chi \chi \to \tau^+ \tau^-$\footnote{As the DM is leptophilic, we only have $\chi \chi \to e^{+} e^{-}, \mu^{+} \mu^{-},~ \text{and}~\tau^{+} \tau^{-}$. From the Fermi-LAT experiment, $\chi \chi \to \tau^{+} \tau^{-}$ annihilation channel has the most stringent bound on $\vevof{\sigma v}$ bound compared to other channels. Therefore, we only focus on this channel in our study.}. In Fig.~\ref{fig:id}, we illustrate the relic allowed parameter space in $\vevof{\sigma v}_{\chi \chi \rightarrow \tau^+ \tau^-}$ versus $m_{\chi}$ plane. We find that DM mass $\lesssim$ 30 (25) GeV is ruled out from indirect detection bounds put by Fermi-LAT data for DMEFT (DMEFT + $\nu$DMEFT) scenario.

\section{Collider Phenomenology of Leptophilic DM}
\label{sec3}
\begin{figure}[htb!]
	\begin{tikzpicture}
		\begin{feynman}
			\vertex [dot, ultra thick] (blob) at (0,0) {};
			\vertex (b)  at (-1.75,1);
			\vertex (c) at (-0.3,2.5) {\Large $\gamma$};
			\vertex (g1)  at (-3.5,2) {\Large $e^-$};
			\vertex (g2) at (-3.5,-2)  {\Large $e^+$};
			\vertex (t2) at (3.5,2)  {\Large $\chi$};
			\vertex (t1) at (3.5,-2)   {\Large $\bar{\chi}$};
			
			\diagram* {
				(g1) -- [fermion, ultra thick, arrow size=2.2pt] (b) -- [anti fermion, ultra thick, arrow size=2.2pt] (blob) -- [fermion, ultra thick, arrow size=2.2pt] (g2),
				(c) -- [photon, ultra thick] (b),
				(t1) -- [fermion, ultra thick, arrow size=2.2pt] (blob) -- [fermion, ultra thick, arrow size=2.2pt] (t2),
			};
			\draw[fill=black] (blob) circle(3mm);
		\end{feynman} 
	\end{tikzpicture}
	\begin{tikzpicture}
		\begin{feynman}
			\vertex (a) at (0,0) ;
			\vertex (b)  at (-1.75,-1);
			\vertex (c) at (-0.3,-2) {\Large $\gamma$};
			\vertex (g1)  at (-3.5,2) {\Large $e^-$};
			\vertex (g2) at (-3.5,-2)  {\Large $e^+$};
			\vertex (t2) at (3.5,2)  {\Large $\chi$};
			\vertex (t1) at (3.5,-2)   {\Large $\bar{\chi}$};
			
			\diagram* {
				(g1) -- [fermion, ultra thick, arrow size=2.2pt]  (blob) -- [anti fermion, ultra thick, arrow size=2.2pt] (b)  -- [fermion, ultra thick, arrow size=2.2pt] (g2),
				(c) -- [photon, ultra thick] (b),
				(t1) -- [fermion, ultra thick, arrow size=2.2pt]  (blob) -- [fermion, ultra thick, arrow size=2.2pt] (t2),
			};
			\draw[fill=black] (a) circle(3mm);
		\end{feynman}
	\end{tikzpicture}
	\caption{Feynmann diagram of mono-$\gamma$+ missing energy ($\slashed{E}$) for DM signal at the the $e^+e^-$ colliders.}
	\label{fig:monoph}
\end{figure}
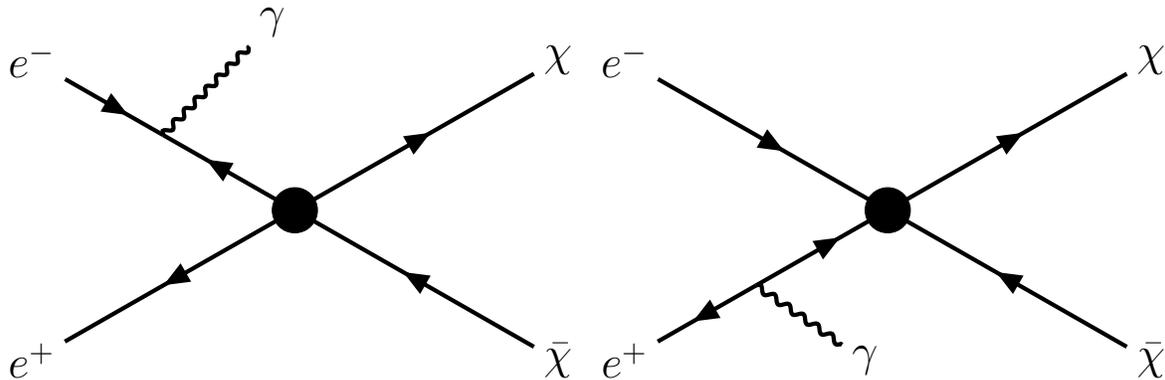

When DM is produced at the collider experiments, it does not interact with the detector. As a result, it goes undetected, effectively disappearing from the measurement apparatus manifesting as ``missing energy'' in the detector's readings. The presence of this missing energy can thus be an indirect signal of dark matter production at the colliders. We consider the pair production of DM particle with an initial-state-radiation (ISR) photon that gives rise to mono-$\gamma$ + missing energy ($\slashed{E}$) final signal, a popular final state to test DM at lepton colliders (Fig.~\ref{fig:monoph}). Missing energy carried away by DM particles is unregistered in the detector while the photon energy ($E_{\gamma}$) can be observed. Due to the absence of the QCD background, the interaction can take place in a much cleaner background at the lepton colliders compared to hadron colliders. Therefore, we study the prospect of probing leptophilic DM at future lepton colliders such as electron-position colliders (International linear collider (ILC) \cite{Behnke:2013xla} with polarised beam) and high energy muon collider ($\mu$C) \cite{Black:2022cth}. 

\begin{figure}
	\begin{tikzpicture}
		\begin{feynman}
			\vertex (a) at (0, 1.9) {\Large $e^-$};
			\vertex (b) at (0, -1.9) {\Large $e^+$};
			\vertex (c) at (2, 0);
			\vertex (c1) at (5, 0);
			\vertex (d) at (7, 1.9) {\Large $\nu$};
			\vertex (e) at (7, -1.9) {\Large $\bar{\nu}$};
			
			\diagram* {
				(a) -- [fermion, ultra thick, arrow size=2.2pt] (c) -- [fermion, ultra thick, arrow size=2.2pt] (b),
				(e) -- [fermion, ultra thick, arrow size=2.2pt] (c1) -- [fermion, ultra thick, arrow size=2.2pt] (d),
				(c) -- [boson, ultra thick, edge label=$Z$] (c1)
			};
		\end{feynman}
	\end{tikzpicture}~~
	\begin{tikzpicture}
		\begin{feynman}
			\vertex (a) at (-1, 1.8) {\Large $e^-$};
			\vertex (b) at (-1, -1.8) {\Large $e^+$};
			\vertex (c) at (2, 1.8);
			\vertex (d) at (2, -1.8);
			\vertex (e) at (5, 1.8) {\Large $\nu$};
			\vertex (f) at (5, -1.8) {\Large $\bar{\nu}$};
			
			\diagram* {
				(a) -- [fermion, ultra thick, arrow size=2.2pt] (c) -- [fermion, ultra thick, arrow size=2.2pt] (e),
				(b) -- [anti fermion, ultra thick, arrow size=2.2pt] (d) -- [anti fermion, ultra thick, arrow size=2.2pt] (f),
				(c) -- [boson, ultra thick, edge label=$W$] (d)
			};
		\end{feynman}
	\end{tikzpicture}
	\begin{tikzpicture}
		\begin{feynman}
			\vertex (f) at (0,0);
			\vertex (g1)  at (-3.5,2) {\Large $e^-$};
			\vertex (g2) at (-3.5,-2)  {\Large $e^+$};
			\vertex (t2) at (3.5,2)  {\Large $\nu_R$};
			\vertex (t1) at (3.5,-2)   {\Large $\bar{\nu}_R$};
			
			\diagram* {
				(g1) -- [fermion, ultra thick, arrow size=2.2pt] (f) -- [anti fermion, ultra thick, arrow size=2.2pt] (blob) -- [fermion, ultra thick, arrow size=2.2pt] (g2),
				(t1) -- [fermion, ultra thick, arrow size=2.2pt] (blob) -- [fermion, ultra thick, arrow size=2.2pt] (t2),
			};
			\draw[fill=black] (blob) circle(3mm);
		\end{feynman} 
	\end{tikzpicture}
	\caption{Top: Feynman diagrams of SM backgrounds contributing to mono-$\gamma$+ missing energy ($\slashed{E}$) for DM signal at the the $e^+e^-$ colliders. For $Z$ mediation, a photon can be radiated from the initial state, contributing to a mono-$\gamma$ final state, whereas, for $W$ mediation, a photon can be radiated not only from the initial state but also from the $W$-boson, contributing to the given final state; bottom: Feynman diagram of non-standard background contribution ($\mathcal{O}_{\text{SM}} \mathcal{O}_{\nu_R}$ operators) to the mono-$\gamma$ final state. The emergence of ISR photon is similar to Fig.~\ref{fig:monoph}.}
	\label{fig:sm.bkg.monoph}
\end{figure}
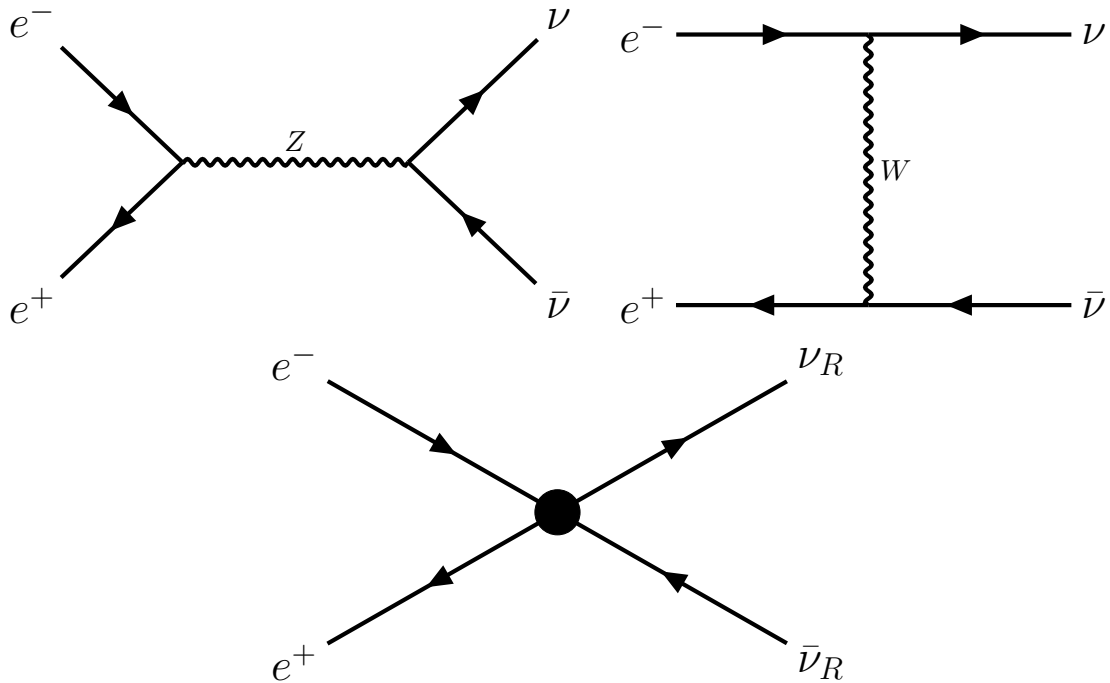

\noindent
The dominant non-interfering SM backgrounds owing to mono-$\gamma$ signal arise from $s$-channel $Z$-mediated and $t$-channel $W$-mediated neutrino pair production at the lepton colliders as shown in the top of Fig.~\ref{fig:sm.bkg.monoph}. If $\mathcal{O}_{\text{SM}} \mathcal{O}_{\nu_R}$ operators are present\footnote{For simplicity we consider only vector operators constructed out of SM leptons and $\nu_R$ ($\mathcal{O}_{\ell \nu}$ and $\mathcal{O}_{L \nu}$).}, then that operator also contributes to the mono-$\gamma$ final state, which is also treated as a non-standard background compared to the DM signal (bottom of Fig.~\ref{fig:sm.bkg.monoph}).  

\begin{figure}[h]
	\centering
	\includegraphics[scale=0.38]{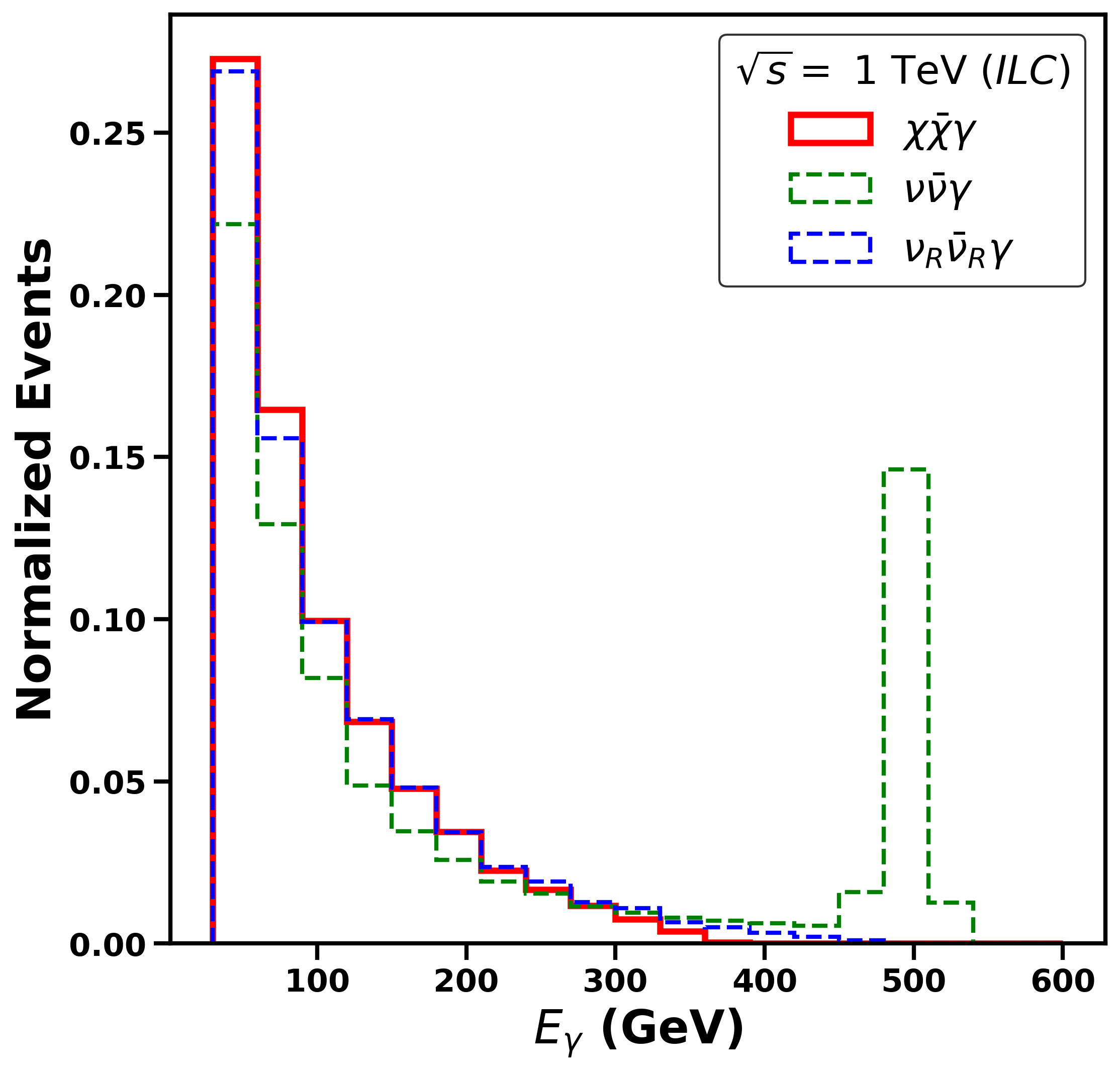} ~~
	\includegraphics[scale=0.38]{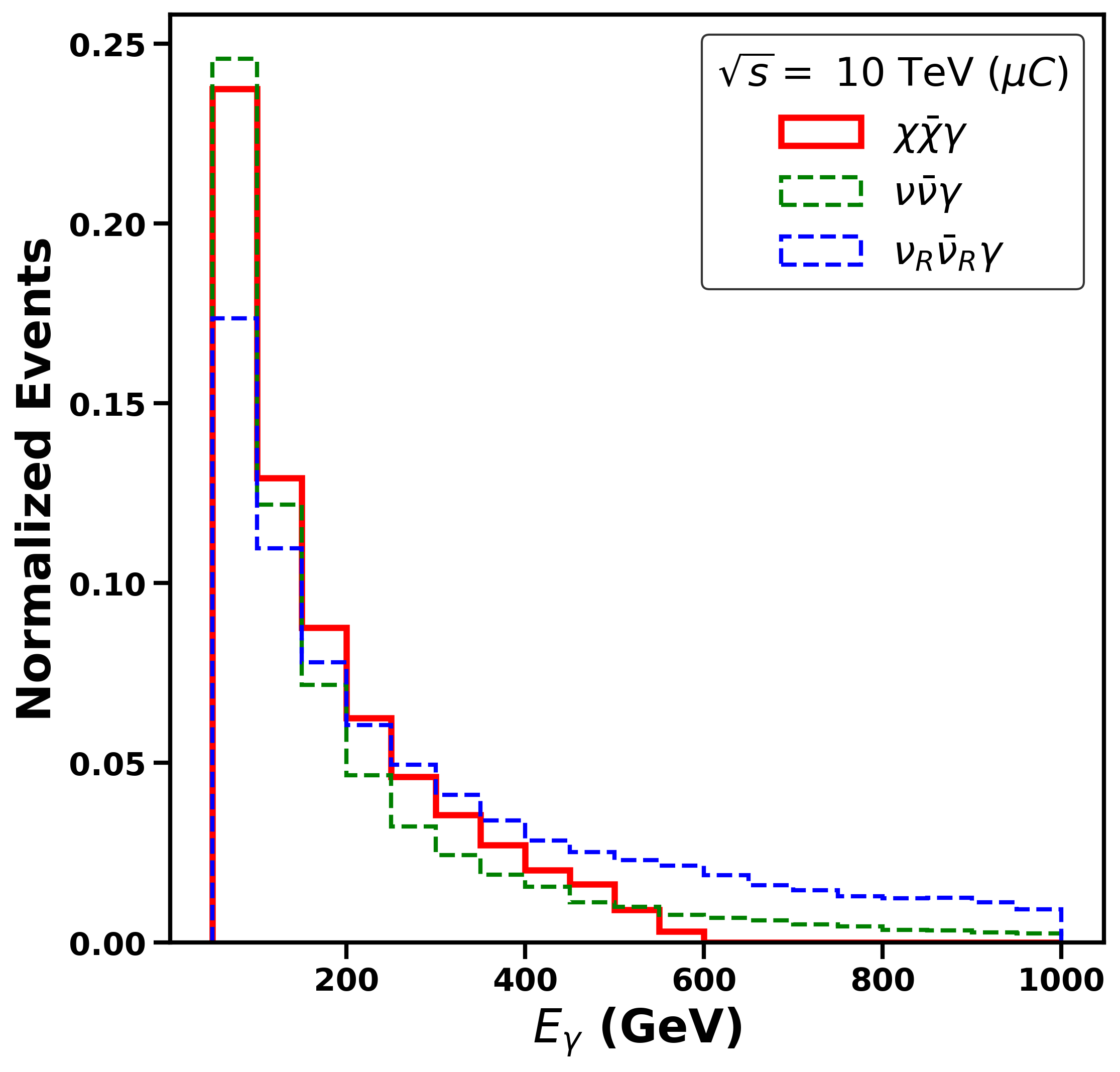}
	
	\caption{Normalised $E_{\gamma}$ distributions for signal and background events at the lepton colliders. Left panel: $\sqrt{s}=1$ TeV (ILC) with DM mass $m_\chi=260$ GeV and cutoff scale $\Lambda = 2450$ GeV, right panel: $\sqrt{s}=10$ TeV ($\mu$C) with DM mass $m_\chi=4.7$ TeV and cutoff scale $\Lambda = 10.5$ TeV.}
	\label{fig:dist}
\end{figure}

We generate signal and background events using {\tt Madgraph} \cite{Alwall:2014hca}, which are then showered and analyzed through {\tt Pythia} \cite{Sjostrand:2014zea}. The detector simulation is performed by {\tt Delphes} \cite{deFavereau:2013fsa}. The UFO file provided to {\tt Madgraph} is created using {\tt FeynRules} \cite{Christensen:2008py}. We apply a cut on the photon's transverse momentum, requiring $P^{\gamma}_T > 5$ GeV, and limit the pseudorapidity ($\eta_{\gamma}$) to $\eta_{\gamma} \leq 2.5$ to ensure that photons are registered in the detector. Additionally, a veto is imposed on jets with $p^{j}_T > 20 \, \text{GeV}$ and leptons with $p^{\ell}_T > 10 \, \text{GeV}$ to secure only one photon is counted. $E_{\gamma}$ is our collider variable to determine the signal-background estimation. Event distributions are shown in Fig.~\ref{fig:dist} for two different types of lepton colliders. We observe a peak at the tail of the SM background spectrum (green dashed line) attributed to the $s$-channel neutrino pair production (left of Fig.~\ref{fig:sm.bkg.monoph}) with an ISR photon. The location of this peak is determined by the mass of the $Z$ boson ($m_Z$) and CM energy $\sqrt{s}$, as described by the following expression
\beq
E^Z_{\gamma}=\frac{1}{2}\sqrt{s}\left(1-\frac{m_Z^2}{s}\right).
\eeq
On the contrary, the signal distribution does not have a peak at the location but has a continuously falling distribution which ends at \cite{Liu:2019ogn}
\beq
E^{\chi}_{\gamma} \le \frac{1}{2}\sqrt{s}\left(1-\frac{4 m_{\chi}^2}{s}\right).
\label{eq:egamma}
\eeq
The difference in the photon energy spectrum allows for the effective removal of a significant portion of the background using
\beq
E_{\gamma} < \bar{E}_{\gamma} = \rm{min}(E^Z_{\Gamma}-5\Gamma_Z,E^{\chi}_{\gamma}),
\eeq
where $\Gamma_z$ is the decay width of $Z$ boson. Reduction of non-interfering SM background enhances the signal significance after employing the above cut on $E_{\gamma}$. As we increase the CM energy of lepton colliders ({\it e.g.,} 10 TeV muon collider), the $s$-channel contribution to the mono-$\gamma$ signal decreases. As a result, we do not have any second peak in the $E_{\gamma}$ distribution at higher CM energy. The low-energy peak close to $E_{\gamma}$ near 0, induced by $W$-mediated $t$-channel diagram, almost overlaps with the signal distribution and could not be effectively diminished. Even though a substantial amount of SM background persists after applying the kinematical cuts, beam polarization plays a very important role in reducing this background while minimally impacting the signal. We present the signal and background cross-sections in Table~\ref{tab:xsec} for different choices of beam polarization. With \{$-20\%,+80\%$\} polarization combination, the $\nu \bar{\nu} \gamma$ background is suppressed by a factor of six while the signal shows a slight increment. Thus, selecting this polarization choice enhances the signal significance. 

\begin{table}[hbt!]
	\centering
	\begin{tabular}{|c|c|c|c|}
		\hline 
		\text{Beam Polarization} & \multicolumn{3}{|c|}{\text{Cross-section} \text{(fb)}} \\
		\cline{1-4}
		$\left\{P_{e^{+}}, P_{e^{-}}\right\}$ & $\chi \bar{\chi} \gamma$ $(S)$ & $\nu \bar{\nu} \gamma$ $(B1)$ & $\nu_R \bar{\nu}_R \gamma$ $(B2)$ \\
		\hline
		\{0\%, 0\%\} & 12.34 & 2450 & 19.65 \\
		\{$+20\%,+80\%$\} & 10.38 & 630 & 16.50 \\
		\{$-20\%,+80\%$\} & 14.30 & 463 & 22.79 \\
		\{$+20\%,-80\%$\} & 14.33 & 5218 & 22.92 \\
		\{$-20\%,-80\%$\} & 10.35 & 3476 & 16.54 \\
		\hline
	\end{tabular}
	\caption{Cross-sections for various beam polarizations at $\sqrt{s} = 1$ TeV and relic satisfied EFT parameters, $m_{\chi} = 260$ GeV and $\Lambda =  2450$ GeV.}
	\label{tab:xsec}
\end{table}

\begin{figure}[h]
	\centering
	\includegraphics[scale = 0.29]{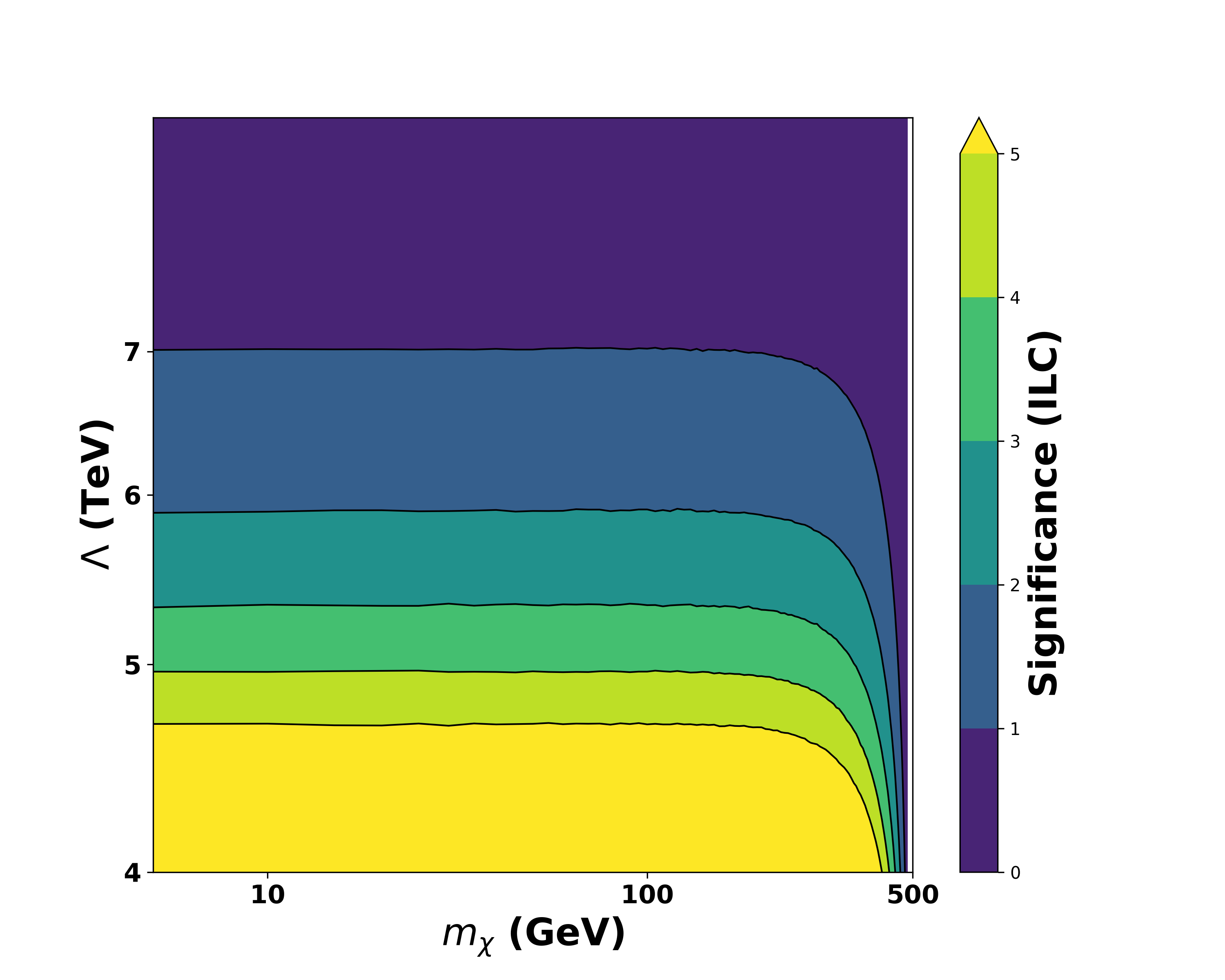}
	\includegraphics[scale = 0.29]{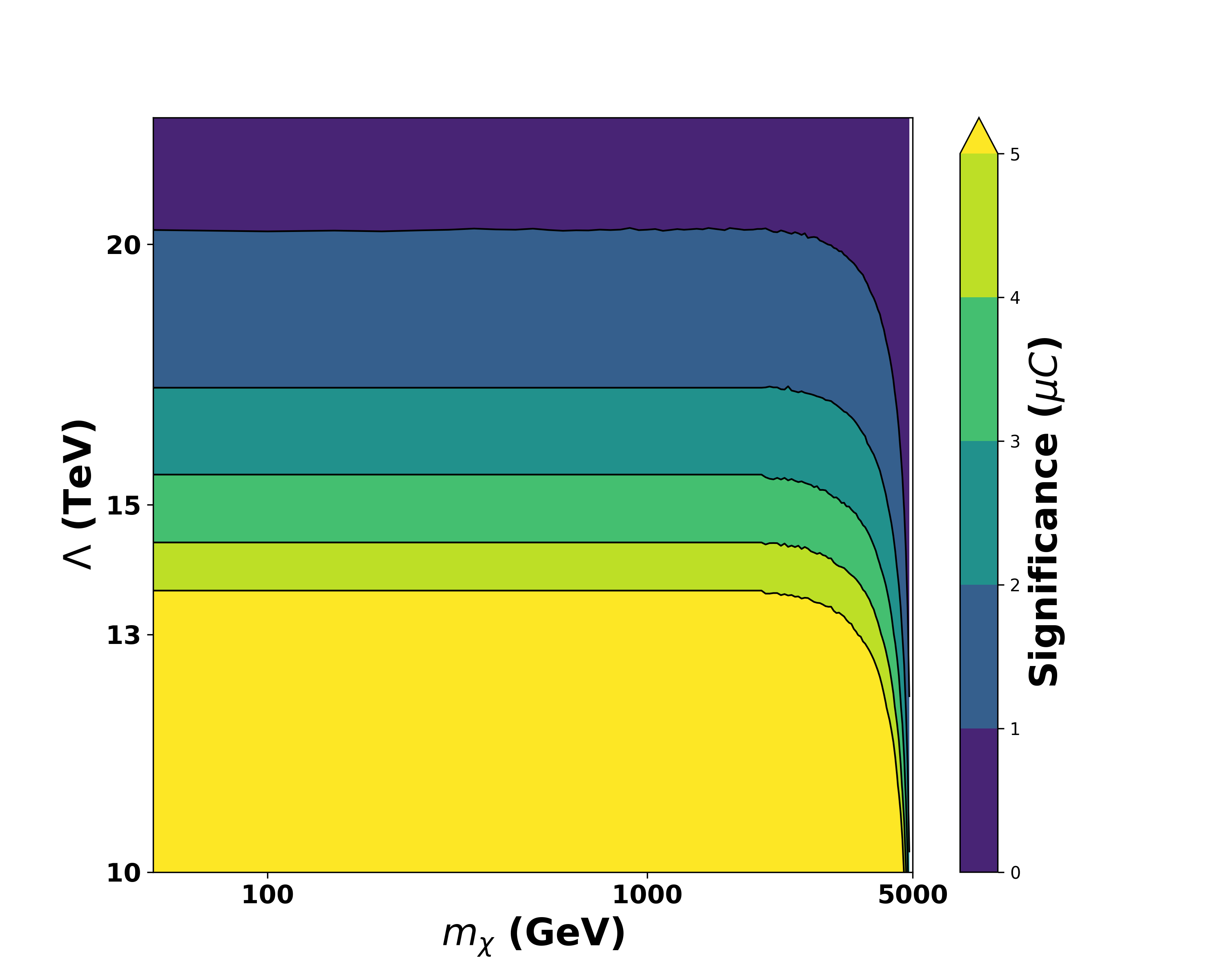}
	
	\caption{Signal significance variations in $\Lambda-m_{\chi}$ plane at the lepton colliders. Left panel: $\sqrt{s}=1$ TeV (ILC), right panel: $\sqrt{s}=10$ TeV ($\mu$C).}
	\label{fig:signi}
\end{figure}

The signal significance ($\mathfrak{Z}$) is defined as
\beq
\mathfrak{Z} = \sqrt{2 \left[(S + B) \log \left(1 + \frac{S}{B} \right) - S \right]},
\eeq
where $S$ and $B$ are the signal and total number of background events, respectively. In the limit of $B \ll S$, $\mathfrak{Z}$ boils down to $\mathfrak{Z} \simeq S/\sqrt{B}\propto 1/\Lambda^4$.
Fig.~\ref{fig:signi} illustrates the signal significance in the $\Lambda$ – $m_{\chi}$ plane using colored bands. The significance decreases with larger $\Lambda$, as the signal cross-section is proportional to $1/\Lambda^4$. Additionally, the $\mathfrak{Z}$ reduces with increasing $m_{\chi}$ due to the limited phase space.

In the above discussion, we have discussed the collider analysis at the ILC and $\mu$C with their maximum reach. Similar analysis can be done at the future circular collider (FCC-ee) \cite{Agapov:2022bhm} and compact linear collider (CLIC) \cite{Brunner:2022usy} as well. Using maximum reach and judicious choice of polarization (where applicable), the 3$\sigma$ sensitivity of $\Lambda$ at the FCC-ee (CLIC) is expected to reach approximately 1.5 (8.8) TeV. We would like to mention that the LEP limits \cite{Assmann:2002th} already rule out $\Lambda$ below approximately 500 GeV. We include these sensitivities and bounds in the final allowed parameter space to be discussed below.

\begin{table}[hbt!]
	\centering
	\begin{center}
		\begin{tabular}{ |c|c|c| } 
			\hline
			Colliders & ($\sqrt{s},\mathfrak{L}_{\text{int}}$) & Beam polarization \\ 
			& & $\{P_{e^+},P_{e^-}\}$\\
			\hline
			FCC-ee & (365 GeV, 340 $\rm{fb^{-1}}$) & $-$ \\ 
			ILC & (1 TeV, 8 $\rm{ab^{-1}}$) & $\{\pm 20\%,\pm 80\%\}$ \\ 
			CLIC & (3 TeV, 5 $\rm{ab^{-1}}$) & $\{\pm 00\%,\pm 80\%\}$ \\ 
			$\mu$C & (10 TeV, 10 $\rm{ab^{-1}}$) & $-$ \\ 
			\hline
		\end{tabular}
		\caption{Details of the maximum reach (highest CM energy and integrated luminosity ($\mathfrak{L}_{\text{int}}$)) of different lepton colliders.}
		\label{tab:lep.col}
	\end{center}
\end{table}

\section{Cosmological signatures}
\label{sec4}
The presence of three species of $\nu_R$ can have interesting cosmological signatures. The $\nu_R$'s being light, can contribute to the radiation energy density of the early Universe.  Light Dirac neutrino models often lead to enhanced $\Delta {\rm N}_{\rm eff}$, some recent works on which can be found in \cite{Abazajian:2019oqj, FileviezPerez:2019cyn, Nanda:2019nqy, Han:2020oet, Luo:2020sho, Borah:2020boy, Adshead:2020ekg, Luo:2020fdt, Mahanta:2021plx, Du:2021idh, Biswas:2021kio, Borah:2022obi, Borah:2022qln, Li:2022yna, Biswas:2022fga, Adshead:2022ovo, Borah:2023dhk, Borah:2022enh, Das:2023oph, Esseili:2023ldf, Das:2023yhv}. After $e^{\pm}$ annihilation, the radiation energy density of the universe in presence of 3 $\nu_R$s can be written as 
\begin{equation}
	\rho_{r} = \rho_{\gamma} + \rho_{\nu_{L}} + \rho_{\nu_{R}} = \left(1+\frac{7}{8}\left(\frac{4}{11}\right)^{4/3} N_{\rm eff}\right) \rho_{\gamma},
\end{equation}
where $N_{\rm eff}$ represents the effective number of relativistic species. For SM only and under the assumption of instantaneous decoupling of three SM neutrinos, $N_{\rm eff}$ takes value of $3$. Accounting the non-instantaneous decoupling of SM neutrinos along with flavour oscillations and plasma correction of quantum electrodynamics, the SM $N_{\rm eff}$ value is obtained as $N_{\rm eff}^{\rm SM} = 3.045$ \cite{Mangano:2005cc, Grohs:2015tfy,deSalas:2016ztq}. A deviation from $N_{\rm eff}^{\rm SM}$ indicates presence of BSM physics. The current bound on $N_{\rm eff}$ from Planck 2018 data is given as $N_{\rm eff}=2.99^{+0.34}_{-0.33}$ at $2\sigma$ CL including baryon acoustic oscillation (BAO) data. The translated bound on $\Delta N_{\rm eff}$ at $2\sigma$ can be written as $\Delta N_{\rm eff} \lesssim 0.285$. The latest DESI 2024 data give a slightly weaker bound $\Delta N_{\rm eff} \lesssim 0.4$ at $2\sigma$ CL \cite{DESI:2024mwx}. A similar bound also exists from big bang nucleosynthesis (BBN) $2.3 < {\rm N}_{\rm eff} <3.4$ at $95\%$ CL \cite{Cyburt:2015mya}. Future CMB experiment CMB Stage IV (CMB-S4) is expected reach a much better sensitivity of $\Delta {\rm N}_{\rm eff}={\rm N}_{\rm eff}-{\rm N}^{\rm SM}_{\rm eff}
= 0.06$ \cite{Abazajian:2019eic}, taking it closer to the SM prediction. Another future experiment CMB-HD \cite{CMB-HD:2022bsz} can probe $\Delta N_{\rm eff}$ upto $0.014$ at $1\sigma$.

The change in $N_{\rm eff}$ due to presence of $\nu_R$s can be parameterised as
\begin{equation}
	\Delta N_{\rm eff} = N_{\rm eff} - N_{\rm eff}^{\rm SM} = \frac{\rho_{\nu_{R}}}{\rho_{\nu_{L},1}}, 
\end{equation}
where $\rho_{\nu_{L},1}$ denotes energy density of a single SM neutrino species. If $\nu_R$s are in thermal equilibrium at an initial temperature and decouple at a later temperature $T_{\rm dec}$, their contribution to $\Delta N_{\rm eff}$ can be written as 
\begin{equation}
	\Delta N_{\rm eff} = 0.047 \times 3 \left(\frac{106.75}{g_{*\rho}(T_{\rm dec})}\right)^{4/3}.
	\label{eq:dneff}
\end{equation}
Here $g_{*\rho}(T_{\rm dec})$ denotes sum of all relativistic SM degrees of freedom at the decoupling temperature. We consider different operators listed in Table \ref{tab:rhn.ops} one by one and study their contributions to $N_{\rm eff}$. The details can be found in Appendix \ref{appen1}. In Fig. \ref{fig:Neff_combined}, we show the combined results for these cases. The details of these cases are summarised in Table \ref{table_Neff}. For cases 1 to 5, the contribution of $\nu_{R}$ energy density to $\Delta N_{\rm eff}$ do not depend on $\chi$ mass, assuming the latter to be a typical WIMP having mass above the GeV scale. However, for case 6, the contribution to $\Delta N_{\rm eff}$ strongly depends on $m_{\chi}$ as $\nu_R$ thermalises only through its interactions with DM. In Fig. \ref{fig:Neff_combined}, the green line for case 6 is obtained by considering different $m_{\chi}$ for different $\Lambda$ such that correct DM abundance is satisfied. Clearly, case 1 faces the tightest constraints whereas case 6 is the least constrained. Variation of $N_{\rm eff}$ for different operators can be understood simply by considering the corresponding decoupling temperature $T_{\rm dec}$ of $\nu_R$. If $\nu_R$ couples to all fermions in the SM, the decoupling occurs late, leading to smaller $g_{*\rho}(T_{\rm dec})$ and hence larger $\Delta N_{\rm eff}$ following Eq. \eqref{eq:dneff}. On the other hand, if $\nu_R$ couples only to the third generation quarks (as in case 5), the $\nu_R$ decoupling occurs much earlier due to Boltzmann suppression in heavy quark's number densities, leaving a smaller $\Delta N_{\rm eff}$. We choose the operators from Table \ref{tab:rhn.ops} in such a way that the other operators do not lead to bounds outside the range obtained for case 1 and case 6.

\begin{figure}[htb!]
	\centering
	\includegraphics[width=0.45\linewidth]{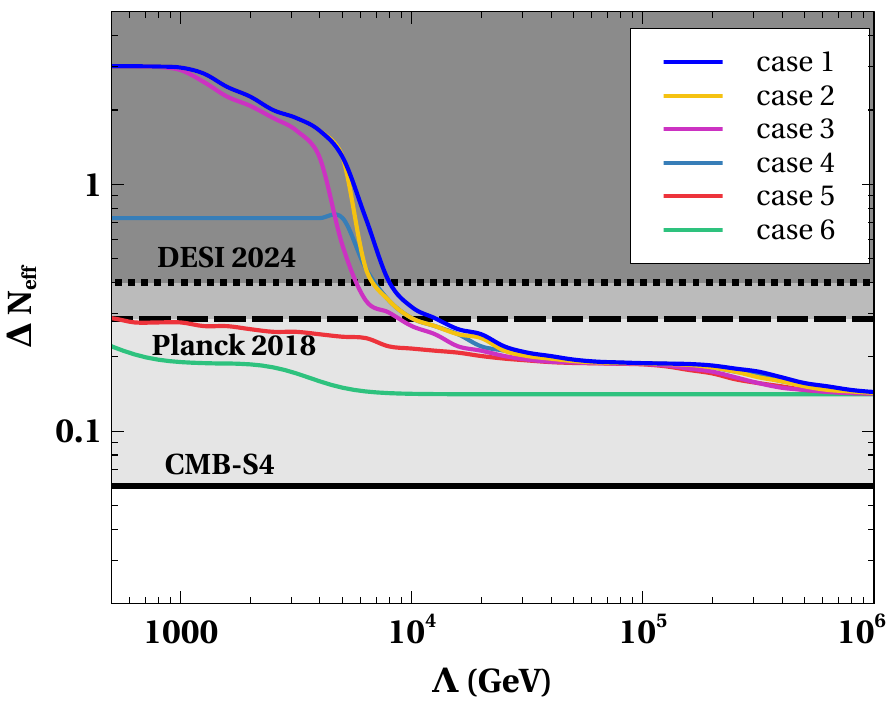}
	\caption{Variation of $\Delta N_{\rm eff}$ with cutoff scale $\Lambda$ for different cases.}
	\label{fig:Neff_combined}
\end{figure}

\begin{table}[h] 
	\centering
	\begin{tabular}{|c|c|c|c|}
		\hline
		\textbf{} & \textbf{Interactions} & \textbf{Disallowed from } & \textbf{Disallowed from } \\ 
		[-6pt]
		\textbf{} & \textbf{($\overline{f}_{L,R} \gamma^\mu \psi_{L,R}$ type only)} & \textbf{Planck 2018} & \textbf{DESI 2024} \\
		\hline
		case 1 & All fermions & $\Lambda \lesssim 11.6 $ TeV  & $\Lambda \lesssim 7.8 $ TeV \\ 
		\hline
		case 2 & leptons & $\Lambda \lesssim  10 $ TeV & $\Lambda \lesssim 6.7 $ TeV \\
		\hline
		case 3 & SM neutrinos & $\Lambda \lesssim 8 $ TeV & $\Lambda \lesssim 5.2 $ TeV \\
		\hline
		case 4 & quarks & $\Lambda \lesssim 9.8 $ TeV & $\Lambda \lesssim 6.5 $ TeV \\
		\hline
		case 5 & 3rd Gen quarks & $\Lambda \lesssim 0.6 $ TeV & $\Lambda \lesssim 0.001 $ TeV \\
		\hline
		case 6 & Dark matter ($\chi$) & $m_{\chi}$ dependent & $m_{\chi}$ dependent \\
		\hline
		
	\end{tabular}
	\caption{Different possibilities of interaction between $\nu_{R}$ and SM species and the corresponding disallowed values of cut-off scale.}
    \label{table_Neff}
\end{table}

\section{UV Completions}
\label{sec4a}
While we stick to effective operators to describe DM interaction with the SM, such operators can be realised in a UV complete theory by incorporating additional degrees of freedom (DOF). If such additional DOF have mass $M$, we can generate effective DM-SM operators at a scale $\mu\ll M$, by integrating out those heavy fields. For example, the dimension 6 operator $ (\overline{\chi}\gamma^\mu \chi)(\overline{L}\gamma_\mu L)/\Lambda^2$ can be generated by extending the model with additional $U(1)$ gauge boson coupling to DM and leptons as vector current. The same UV completion will also generate $ (\overline{\chi}\gamma^\mu \chi)(\overline{e_R}\gamma_\mu e_R)/\Lambda^2$ type operators. The anomaly-free Abelian gauge models based on $L_\alpha-L_\beta$ quantum numbers \cite{He:1990pn,Ma:2001md,Baek:2008nz,Heeck:2011wj,Das:2013jca} can generate such operators connecting Dirac fermion DM and SM leptons. On the other hand, gauged $B-L$ model with Dirac fermion DM having non-zero $B-L$ charge will lead to effective dimension 6 operators relating DM with all SM fermions including $\nu_R$. It is worth mentioning that $\nu_R$'s are automatically part of gauged $B-L$ model due to anomaly cancellation requirements. If we prevent Majorana mass of $\nu_R$ by due to chosen charge or scalar content, one can realise Dirac neutrino in this model \cite{Heeck:2014zfa, Abazajian:2019oqj, Mahanta:2021plx}. Integrating out the $Z'$ also leads to $\nu_R$-SM operators like  $ (\overline{\nu_R}\gamma^\mu \nu_R)(\overline{f}\gamma_\mu f)/\Lambda^2$ with $f$ being an SM fermion. On the other hand, if DM-SM interactions arise due to an Abelian gauge boson, with $\nu_R$'s being uncharged under the corresponding gauge symmetry, we can have DM-SM and DM-$\nu_R$ operators by introducing a scalar field $\phi$ which couples DM and $\nu_R$ like $y \bar{\chi} \nu_R \phi$ \cite{Biswas:2021kio}. Integrating out the heavy singlet scalar $\phi$ leads to DM-$\nu_R$ operators. Similarly, $\nu_R$-SM operators of the type $(\bar{d} \gamma^\mu u) (\bar{\nu}_R \gamma_\mu e)/\Lambda^2$ can be obtained in models with charged vector boson like left-right symmetric model with Dirac neutrinos \cite{Borah:2020boy, Borah:2017leo}. Other operators, shown in table \ref{tab:dm.ops}, \ref{tab:rhn.ops} can also be realised within different UV complete frameworks, which we do not elaborate further.

Here, we briefly discuss the details of one possible UV completion based on anomaly free gauged $L_e-L_\mu$ symmetry. A vector-like fermion $\chi$ charged under this symmetry can naturally lead to dimension 6 effective operators of the type $ (\overline{\chi}\gamma^\mu \chi)(\overline{L_{e, \mu}}\gamma_\mu L_{e, \mu})/\Lambda^2$ at low energy. The field content of the model and the corresponding quantum numbers are given in table \ref{tab:model}. Due to the leptophilic nature of the symmetry, quarks remain uncharged under it. The SM-like Higgs doublet $H_1$ gives masses to all charged fermions, whereas three additional Higgs doublets $H_{2,3,4}$ generate light Dirac neutrino masses with the desired flavour structure. The singlet scalar $\phi$ leads to spontaneous breaking of the $U(1)_{L_e-L_\tau}$ gauge symmetry. As mentioned earlier, global lepton and baryon number conservations are assumed to keep any $L, B$ violating interactions away.
\begin{table}[]
    \centering
    \begin{tabular}{|c|c|c|}
    \hline
    Field & $SU(2) \times U(1)_Y$ & $U(1)_{L_e-L_\mu}$ \\
    \hline
        $L_e, L_\mu, L_\tau$ & $\left ( 2, -\frac{1}{2} \right)$  & $\left ( 1, -1, 0 \right )$\\
        $e_R, \mu_R, \tau_R$ & $\left ( 1, -1 \right)$  & $\left ( 1, -1, 0 \right )$\\
        $\nu^e_R, \nu^\mu_R, \nu^\tau_R$ & $\left ( 1, 0 \right)$  & $\left ( 1, -1, 0 \right )$\\
        $\chi$ & $\left ( 1, 0 \right)$  & $n_\chi$\\
        $H_1, H_2, H_3, H_4$ & $\left ( 2, \frac{1}{2} \right)$ & $\left ( 0, -2, -1, 1 \right)$ \\
        $\phi$ & $\left ( 1, 0 \right)$ & 1 \\
         \hline
    \end{tabular}
    \caption{Relevant field content of one possible UV completion based on $L_e-L_\mu$ gauge symmetry.}
    \label{tab:model}
\end{table}
The relevant part of the Yukawa Lagrangian is given by
\begin{align}
    -\mathcal{L}_Y & \supset (Y_{e} \overline{L_e} e_R +Y_{\mu} \overline{L_\mu} \mu_R+Y_{\tau} \overline{L_\tau} \tau_R) \tilde{H_1} + y_{ee} \overline{L_e} \tilde{H_1} \nu^e_R+ y_{e\mu} \overline{L_e} \tilde{H_2} \nu^\mu_R + y_{e\tau} \overline{L_e} \tilde{H_3} \nu^\tau_R \nonumber \\
    & + y_{\mu \mu} \overline{L_\mu} \tilde{H_1} \nu^\mu_R + y_{\mu \tau} \overline{L_e} \tilde{H_4} \nu^\tau_R + y_{\tau \tau} \overline{L_\tau} \tilde{H_1} \nu^\tau_R + y_{\tau e} \overline{L_\tau} \tilde{H_4} \nu^e_R + y_{\tau \mu} \overline{L_\tau} \tilde{H_3} \nu^\mu_R +{\rm h.c.}
    \end{align}
where Majorana terms are forbidden by the global lepton number symmetry. While the model predicts diagonal charged lepton mass, the light Dirac neutrino mass matrix can be written as 
\begin{equation}
    m_\nu = \begin{pmatrix}
        y_{ee} \langle H_1 \rangle & y_{e\mu} \langle H_2 \rangle & y_{e\tau} \langle H_3 \rangle \\
        0 & y_{\mu \mu} \langle H_1 \rangle & y_{\mu \tau} \langle H_4 \rangle \\
        y_{\tau e} \langle H_4 \rangle & y_{\tau \mu} \langle H_3 \rangle & y_{\tau \tau} \langle H_1 \rangle
    \end{pmatrix}
\end{equation}
where $\langle H_i \rangle$ denotes the vacuum expectation value (VEV) of the neutral component of $H_i$. The one-zero texture in the Dirac neutrino mass matrix is allowed by neutrino oscillation data \cite{Borgohain:2020csn}. In fact, even if we have a three Higgs doublet setup instead of four Higgs doublets, corresponding neutrino mass matrix with two or three texture-zeros are allowed by neutrino oscillation data \cite{Borgohain:2020csn}.

We can get rid of the requirement of multiple Higgs doublets and fine-tuned Dirac Yukawa couplings (particularly the diagonal ones) by considering a seesaw realisation. For example, type-I Dirac seesaw \cite{Borah:2017dmk} will involve the introduction of three vector-like neutral fermion singlets $N_{1,2,3}$ with $L_{e}-L_\mu$ charge $1, -1, 0$ respectively. An additional scalar singlet $\eta$ with zero $L_{e}-L_\mu$ is required. Direct coupling of lepton doublet and $\nu_R$ can be prevented by a discrete $Z_2$ symmetry under which $\nu_R, \eta$ are odd while all other fields are even. Two scalar singlets $\phi_{1,2}$ with $L_{e}-L_\mu$ charge $1,2$ respectively are required to break the gauge symmetry spontaneously while generating the desired neutrino mass and mixing. The Lagrangian relevant for type-I Dirac seesaw is
\begin{align}
    -\mathcal{L}_Y & \supset (y_{ee} \overline{L_e} N_{1R} +y_{\mu \mu} \overline{L_\mu} N_{2R}+y_{\tau \tau} \overline{L_\tau} N_{3R}) \tilde{H_1} + (y'_{ee} \overline{N_{1L}} \nu^e_{R} +y'_{\mu \mu} \overline{N_{2L}} \nu^\mu_{R}+y'_{\tau \tau} \overline{N_{3L}} \nu^\tau_{R}) \eta \nonumber \\
    & +M_{ii} \overline{N_{iL}} N_{iR} + Y_{12} \overline{N_1} N_2 \phi^\dagger_2 + Y_{13} \overline{N_1} N_3 \phi^\dagger_1 +  Y_{21} \overline{N_2} N_1 \phi_2 + Y_{23} \overline{N_2} N_3 \phi_1 + Y_{31} \overline{N_3} N_1 \phi_1 \nonumber \\
    & + Y_{32} \overline{N_3} N_2 \phi^\dagger_1+{\rm h.c.}
\end{align}
After the scalar fields $\eta, \phi_{1,2}$ and the SM Higgs $H_1$ acquire non-zero VEV, the light neutrino mass arises from the type-I Dirac seesaw given by \cite{Borah:2017dmk}
\begin{equation}
    m_{\nu} = -m_D M^{-1}_N m'_D
\end{equation}
where $m_D = {\rm diag}(y_{ee}, y_{\mu \mu}, y_{\tau \tau}) \langle H_1 \rangle, m'_D = {\rm diag}(y'_{ee}, y'_{\mu \mu}, y'_{\tau \tau}) \langle \eta \rangle $ and the heavy singlet fermion mass matrix is 
\begin{equation}
    M_N = \begin{pmatrix}
        M_{11} & Y_{12} \langle \phi_2 \rangle  & Y_{13} \langle \phi_1 \rangle  \\
        Y_{21} \langle \phi_2 \rangle & M_{22} & Y_{23} \langle \phi_1 \rangle \\
        Y_{31} \langle \phi_1 \rangle & Y_{32} \langle \phi_1 \rangle & M_{33}
        \end{pmatrix}.
\end{equation}
Due to the diagonal nature of $m_D, m'_D$, the desired light neutrino mixing is effectively generated by the flavour structure of $M_N$. In both the examples of neutrino mass generation with $L_e-L_\mu$ gauge symmetry, the DM is a vector-like fermion charged under this gauge symmetry and has a bare mass term $m_\chi \overline{\chi}\chi$. At a scale much below $U(1)_{L_e-L_\mu}$ gauge symmetry breaking, one can generate DM-SM as well as $\nu_R$-SM, $\nu_R$-DM operators of the types discussed in our analysis.

While global lepton number symmetry ensures pure Dirac nature of light neutrinos, it is possible to have small breaking of such global symmetries leading to the appearance of tiny Majorana masses. One possible way in which such global symmetries can get broken is via quantum gravity effects. As gravity does not respect any global symmetries \cite{Kallosh:1995hi}, we can consider higher dimensional $L$-violating operators suppressed by Planck scale $M_{\rm Pl}$. For tiny Majorana mass compared to Dirac mass, neutrinos can be considered as pseudo-Dirac. Depending upon the size of Majorana mass, such pseudo-Dirac neutrinos can have very interesting consequences in neutrino oscillation experiments \cite{Anamiati:2019maf} and the observations of solar neutrinos \cite{Giunti:1992hk}, supernova neutrinos \cite{DeGouvea:2020ang}, astrophysical neutrinos \cite{Beacom:2003eu, Dev:2024yrg} as well as cosmic relic neutrinos \cite{Esmaili:2009fk}. In our EFT setup, we can consider such tiny Majorana masses arising from Planck suppressed operators of the type $\overline{\nu^c_R}\nu_R H^\dagger H/M_{\rm Pl}$. This leads to a new mass splitting $\delta m^2 \lesssim 10^{-16} \, {\rm eV}^2$, much lower than the usual solar or atmospheric mass scales. While such a mass splitting can have consequences for astrophysical neutrinos \cite{Beacom:2003eu}, our analysis related to $N_{\rm eff}$, DM, and collider signatures do not change significantly.

\begin{figure}
	\centering
	\includegraphics[width=0.45\linewidth]{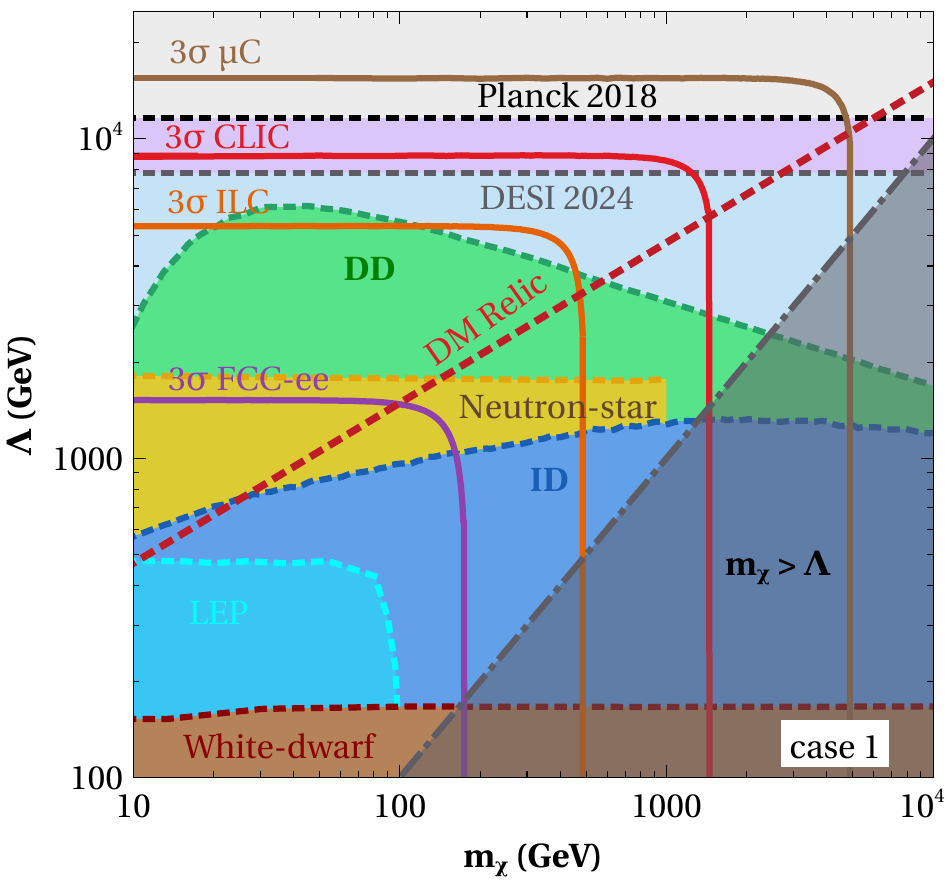}
	\includegraphics[width=0.45\linewidth]{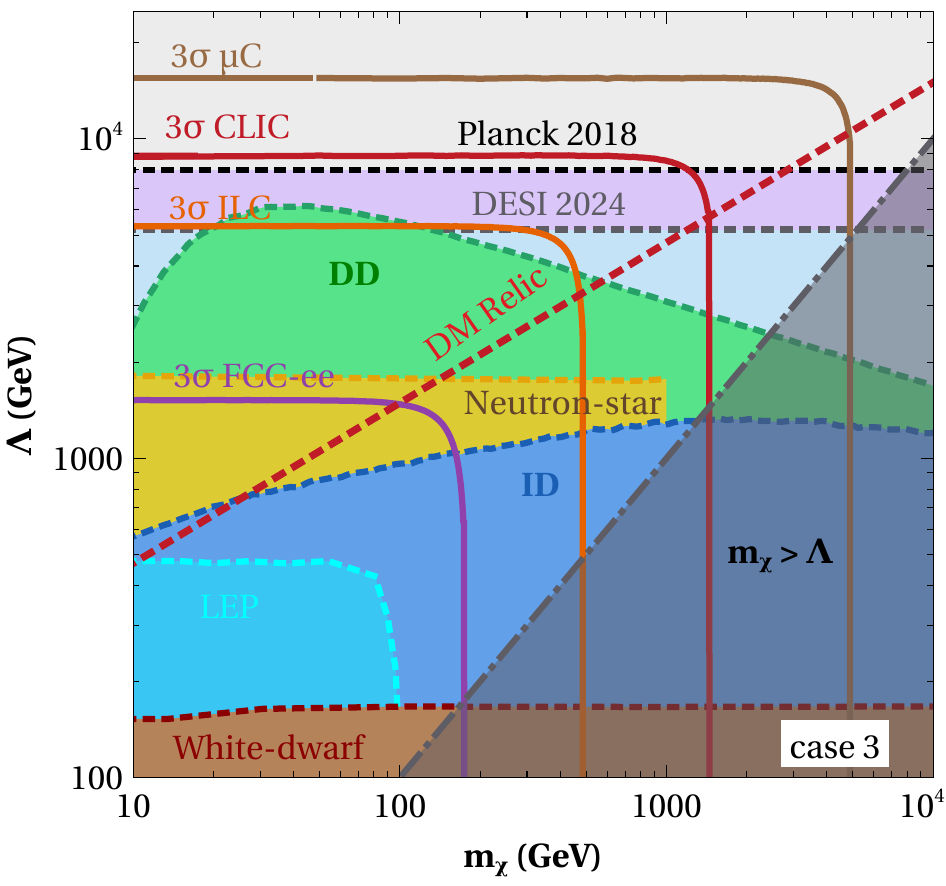}
	\includegraphics[width=0.45\linewidth]{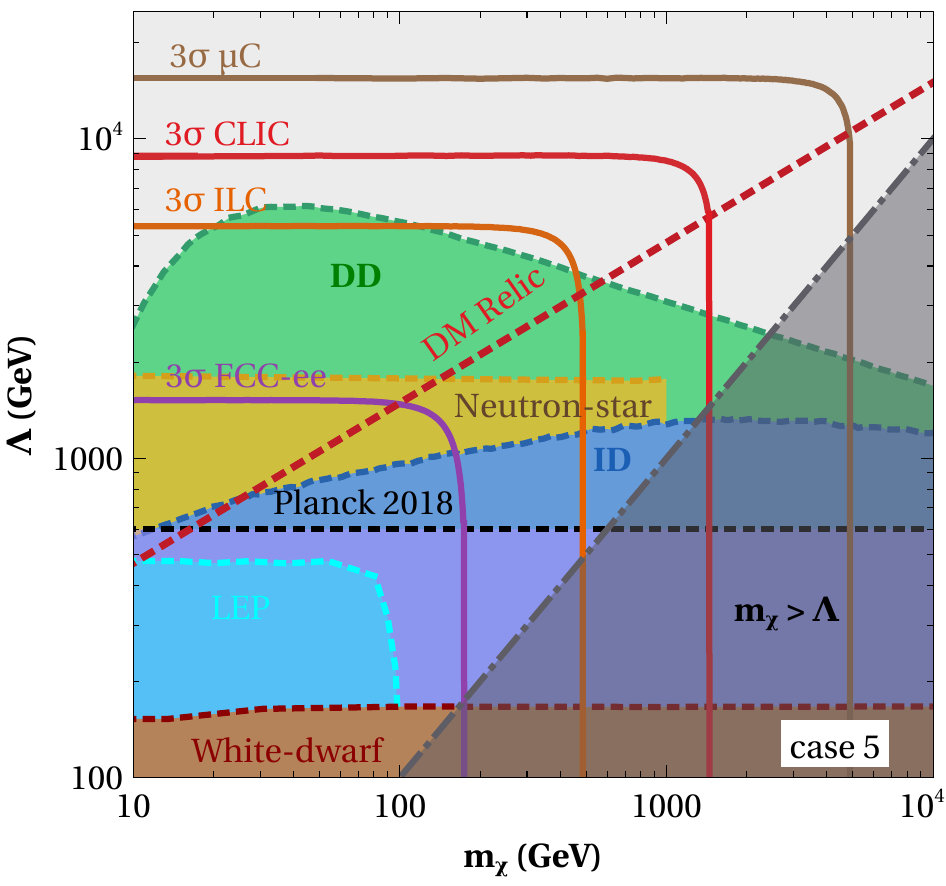}
	\includegraphics[width=0.45\linewidth]{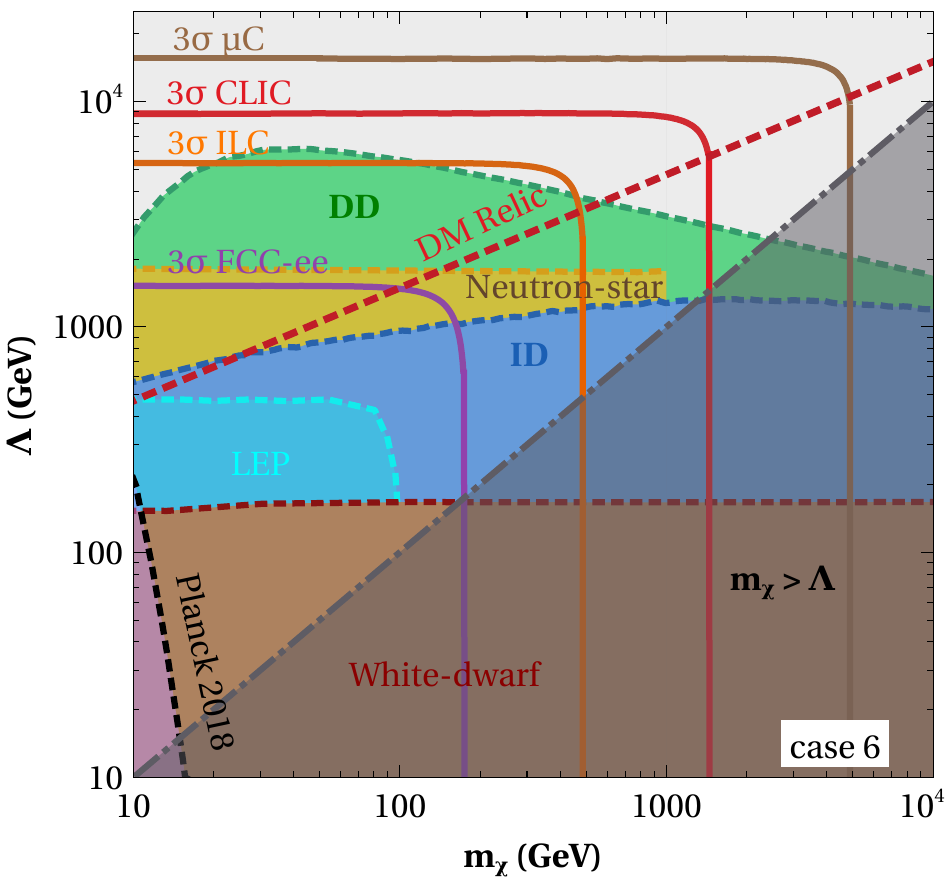}
	\caption{Summary for case 1 (top left), case 3 (top right), case 5 (bottom left), and case 6 (bottom right) showing the parameter space in $\Lambda-m_\chi$ plane. The shaded regions correspond to parameter space disfavoured by different bounds (see text for details). The solid contours correspond to future lepton collider sensitivities while the red dashed contour shows DM relic satisfying region. The entire currently allowed parameter space remains within reach of future CMB experiments like CMB-S4 and CMB-HD.}
	\label{fig:summary1}
\end{figure}





\section{Results and Conclusion}
\label{sec5}
We have studied the discovery prospect of leptophilic dark matter in a scenario where light neutrinos in the SM are of Dirac type. Considering DM, a Dirac fermion, and three copies of right-handed neutrinos $\nu_R$, all singlets under the SM gauge symmetry to be the new fields beyond the standard model, we adopt an EFT approach to study the phenomenology of leptophilic DM, enhanced relativistic degrees of freedom $N_{\rm eff}$ and future discovery prospects at lepton colliders and CMB experiments. While there lies an interesting discovery prospect of leptophilic DM in future lepton colliders for DM mass upto a few TeV and cutoff scale up to a few tens of TeV, current CMB bounds and future prospects crucially depend upon the type of effective operators connecting SM or DM with right handed neutrinos, as summarised in Table \ref{table_Neff}. We summarise our results in $\Lambda-m_\chi$ plane for case 1, case 3, case 5, and case 6 in Fig. \ref{fig:summary1}. In the figure, the current constraints, the future sensitivities, and the theoretical bound are represented by solid, dashed, and dot-dashed lines. Green and blue shaded regions correspond to direct and indirect detection constraints from LUX-ZEPLIN \cite{LZ:2024} and Fermi-LAT \cite{Fermi-LAT:2016afa}, respectively. Yellow and brown shaded regions correspond to astrophysical bounds from white dwarf and neutron star heating, respectively, due to the capture of DM \cite{Bell:2019pyc, Bell:2021fye}. The cyan-shaded region is ruled out by LEP constraints \cite{Assmann:2002th}. The gray shaded region $(m_\chi > \Lambda)$ in the bottom left part corresponds to the parameter space where the EFT description is not valid. The $2\sigma$ constraints on $\Delta N_{\rm eff}$ from Planck 2018 \cite{Planck:2018vyg} and DESI 2024 \cite{DESI:2024mwx} data are shown by black and gray dashed lines, respectively. While they remain horizontal contours for all the cases, it gets $m_\chi$ dependent for case 6 (bottom right panel of Fig. \ref{fig:summary1}) with DESI 2024 contour being outside the plane. The solid contours correspond to $3\sigma$ exclusion limit at the future lepton colliders like muon collider (brown contour) \cite{Black:2022cth}, CLIC (red contour) \cite{Brunner:2022usy}, ILC (orange contour) \cite{Behnke:2013xla} and FCC-ee (purple contour) \cite{Agapov:2022bhm}\footnote{Strictly speaking, for each of these colliders, a portion of the parameter space will make the EFT description invalid where the corresponding center of mass energy $\sqrt{s} > \Lambda$.}. The red dashed line indicates the parameter space satisfying DM relic. While $\nu_R$ coupling to all SM fermions rule out all the parameter space within reach of future colliders (\textit{e.g.}, case 1 shown in the top left panel of Fig. \ref{fig:summary1}), restricting $\nu_R$ interactions to only specific SM fermions (\textit{e.g.}, case 3, case 5 shown in the top right, bottom left panels of Fig. \ref{fig:summary1} respectively) or DM alone (\textit{e.g.}, case 6 shown in the bottom right panel of Fig. \ref{fig:summary1}) open up more parameter space which remain within reach of future lepton colliders. While lepton colliders can probe $\Lambda (m_\chi)$ upto a few tens of TeV (a few TeV), the entire parameter space remains within reach of future CMB experiments. This leads to interesting complementarity among future CMB and lepton colliders in probing such leptophilic DM with light Dirac neutrinos. Additionally, such a scenario remains falsifiable via future observations of neutrinoless double beta decay, which rules out the pure Dirac nature of light neutrinos. We have briefly commented upon possible UV completions of our effective operators connecting SM, DM, and $\nu_R$ while elaborating upon one concrete example based on anomaly free $U(1)_{L_e-L_\mu}$ gauge symmetry.

Before ending, we also summarise the key differences between leptophilic DMEFT with $\nu_R$ (or $\nu$DMEFT) and the usual leptophilic DMEFT studied extensively in the literature. The most noticeable distinction is the possibility of enhanced $N_{\rm eff}$ in $\nu$DMEFT, which can be measured by future CMB experiments. Usual DMEFT does not have such enhancement. Another noticeable difference arises in the DM relic allowed parameter space. For example, with a dark matter mass $m_{\chi}=300$ GeV, the scale $\Lambda$ is 2447 GeV in the DMEFT scenario, while in $\nu$DMEFT scenario, $\Lambda$ extends to 2607 GeV. As far as collider analysis is concerned, we also notice some minor differences in the event distribution due to change in relic allowed DM mass for a fixed cutoff scale. For example, with $\Lambda$ ($\sim$ 2296 GeV) in DMEFT, relic allowed DM mass ($m_{\chi}$) is 260 GeV whereas in $\nu$DMEFT relic allowed DM mass is reduced to 227 GeV. Due to this reduction, there is a change in the $E_{\gamma}$ distribution, according to Eq. \eqref{eq:egamma}. For $m_{\chi}=260$ GeV, $E_{\gamma}$ tail ends at 365 GeV, whereas for $m_{\chi}$ = 227 GeV the tail ends at 396 GeV. However, such minor differences do not change the signal significance as can be seen from Fig. \ref{fig:signi}. While the new one-loop contribution to direct detection in $\nu$DMEFT remains suppressed, the indirect detection signatures can change from usual DMEFT due to the difference in relic allowed parameter space and increase in number of final state particles in DM annihilations. Whether such distinctions in collider or indirect detection signatures can actually be verified experimentally requires more scrutiny and is left for future studies.

\acknowledgments
The work of D.B. is supported by the Science and Engineering Research Board (SERB), Government of India grants MTR/2022/000575 and CRG/2022/000603. The work of N.D. is supported by the Ministry of Education, Government of India via the Prime Minister's Research Fellowship (PMRF) December 2021 scheme. S.J. and B.T. acknowledge Dipankar Pradhan and Abhik Sarkar for useful discussions. S.J. thanks Subhaditya Bhattacharya for insightful communications.

\appendix

\section{$\Delta N_{\rm eff}$ for different operators involving $\nu_R$ and SM or DM}
\label{appen1}

\subsubsection{Interactions with leptons, quarks, and DM} 
The interactions of $\nu_{R}$ with DM and SM fermions and DM are given in Table \ref{tab:rhn.ops}. The total interaction rate of $\nu_{R}$s with SM fermion can be written as
\begin{eqnarray}
	\Gamma = \sum_{f} \frac{n^{\rm eq}_{f} n^{\rm eq}_{\Bar{f}}}{n^{\rm eq}_{\nu_{R}}} \langle \sigma v\rangle_{f\Bar{f} \to \nu_{R}\Bar{\nu}_{R}},
\end{eqnarray}
where $f (\Bar{f})$ denotes SM fermions (anti-fermions) and DM. Here we consider interactions of $\nu_{R}$ with {$\nu_{L}, e, \mu, \tau, u, d, c, s, b, t, \chi$}. The decoupling temperature of $\nu_{R}$ can be approximately calculated by comparing the interaction rate with the Hubble expansion rate
\begin{eqnarray}
	\Gamma (T_{\rm dec}) \sim \mathcal{H} (T_{\rm dec}).
\end{eqnarray}

\begin{figure}[hbt!]
	\centering
	\includegraphics[width=0.45\linewidth]{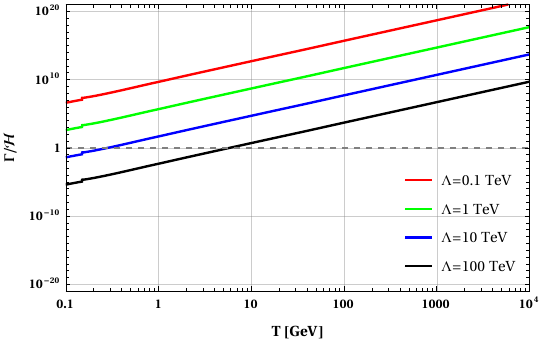}
	\includegraphics[width=0.45\linewidth]{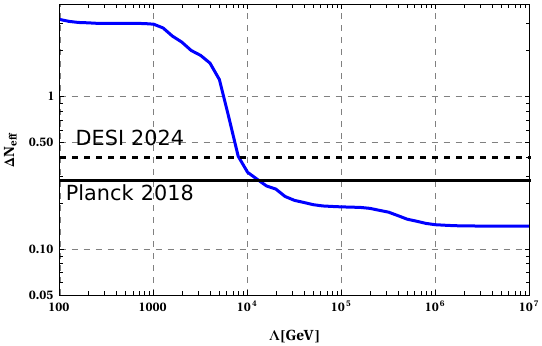}
	\caption{Left panel: comparison of $\nu_R$ interaction rate with Hubble rate of expansion assuming interactions of $\nu_{R}$ with all SM fermions (case 1). Right panel: Variation of $\Delta N_{\rm eff}$ with cutoff scale $\Lambda$ for case 1.}
	\label{fig:Neff_all}
\end{figure}

The left panel of Fig. \ref{fig:Neff_all} shows $\Gamma/\mathcal{H}$ as a function of bath temperature for different values of $\Lambda$. As the Universe cools down below $150$ MeV, quark-gluon plasma makes a transition and forms hadrons. Although $\nu_{R}$ can, in principle, interact with hadrons below $150$ MeV, for the simplicity of calculation, we neglect the interaction of $\nu_{R}$ with hadrons. In the right panel, $\Delta N_{\rm eff}$ as a function of $\Lambda$ is shown. $\Lambda \lesssim 11.6$ TeV gives $\Delta N_{\rm eff}>0.28$, therefore excluded from Planck 2018 data. Similarly, recent results from DESI 2024 data exclude $\Lambda \lesssim 7.8$ TeV.

\subsubsection{Interactions of $\nu_{R}$ with leptons}
\begin{figure}[hbt!]
	\centering
	\includegraphics[width=0.45\linewidth]{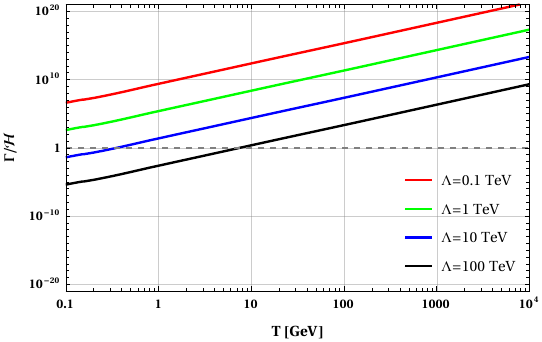}
	\includegraphics[width=0.45\linewidth]{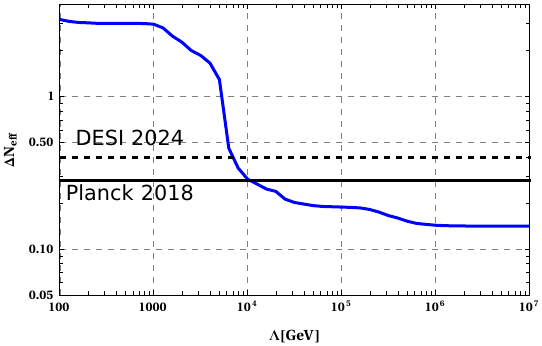}
	\caption{Left panel: comparison of $\nu_R$ interaction rate with Hubble rate of expansion assuming interactions of $\nu_{R}$ with SM leptons only (case 2). Right panel: Variation of $\Delta N_{\rm eff}$ with cutoff scale $\Lambda$ for case 2.}
	\label{fig:Neff_leptons}
\end{figure}

Here we consider the interaction of $\nu_{R}$ only with SM leptons. Due to the absence of interactions with quarks and DM, $\nu_{R}$s decouple from the bath at a slightly higher temperature than the above scenario for a particular value of $\Lambda$. From the constraint on the effective number of relativistic species, $\Lambda \lesssim 10$ TeV ($\Lambda \lesssim 6.7$ TeV) is excluded from Planck 2018 (DESI 2024) data, as shown in Fig. \ref{fig:Neff_leptons}. 

\subsubsection{Interactions of $\nu_{R}$ with $\nu_{L}$}
\begin{figure}[h]
	\centering
	\includegraphics[width=0.45\linewidth]{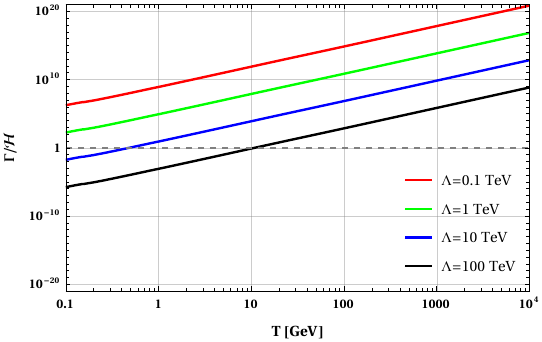}
	\includegraphics[width=0.45\linewidth]{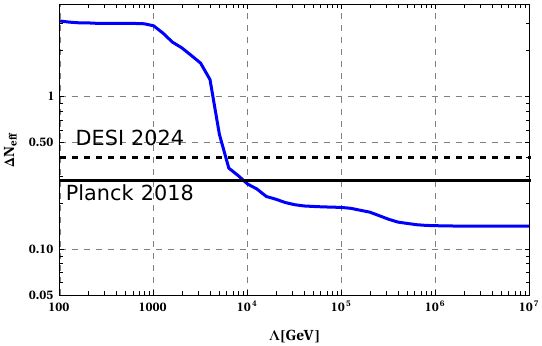}
	\caption{Left panel: comparison of $\nu_R$ interaction rate with Hubble rate of expansion assuming interactions of $\nu_{R}$ with SM neutrinos only (case 3). Right panel: Variation of $\Delta N_{\rm eff}$ with cutoff scale $\Lambda$ for case 3.}
	\label{fig:Neff_nuL}
\end{figure}

From table \ref{tab:rhn.ops}, if we exclude charged leptons from $\mathcal{O}_{L\nu}$ operator, then we find four-Fermi contact interaction, which consists of $\nu_{R}$ and SM neutrinos only. For this scenario, from the constraint on the effective number of relativistic species, $\Lambda \lesssim 8$ ($\lesssim 5.2$) TeV is excluded from Planck 2018 (DESI 2024) data, as shown in Fig. \ref{fig:Neff_nuL}. 

\subsubsection{Interactions of $\nu_{R}$ with quarks}
\begin{figure}[hbt!]
	\centering
	\includegraphics[width=0.45\linewidth]{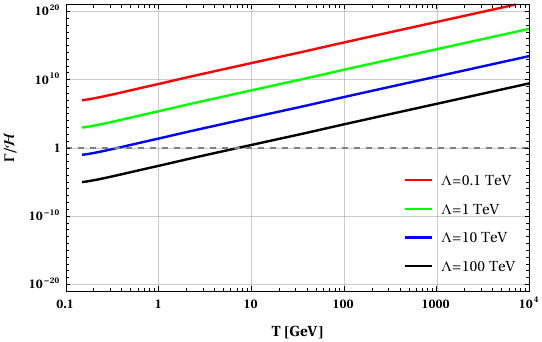}
	\includegraphics[width=0.45\linewidth]{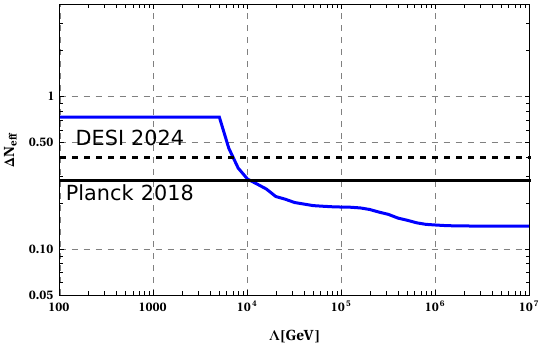}
	\caption{Left panel: comparison of $\nu_R$ interaction rate with Hubble rate of expansion assuming interactions of $\nu_{R}$ with SM quarks only (case 4). Right panel: Variation of $\Delta N_{\rm eff}$ with cutoff scale $\Lambda$ for case 4.}
	\label{fig:Neff_quarks}
\end{figure}

From table \ref{tab:rhn.ops}, the operators relevant for the interactions are $\mathcal{O}_{u\nu}$, $\mathcal{O}_{d\nu}$ and $\mathcal{O}_{Q\nu}$. As mentioned before, below QCD phase transition, we neglect the interaction rate of $\nu_{R}$s with quarks. This gives a constant $\Delta N_{\rm eff}$ at around $0.7$ below $150$ MeV. Here, $\Lambda \lesssim 9.8$ TeV ($\lesssim 6.5$ TeV) is excluded from Planck 2018 (DESI 2024), as shown in Fig. \ref{fig:Neff_quarks}.

\subsubsection{Interactions of $\nu_{R}$ with 3rd generation of quark}
\begin{figure}[hbt!]
	\centering
	\includegraphics[width=0.45\linewidth]{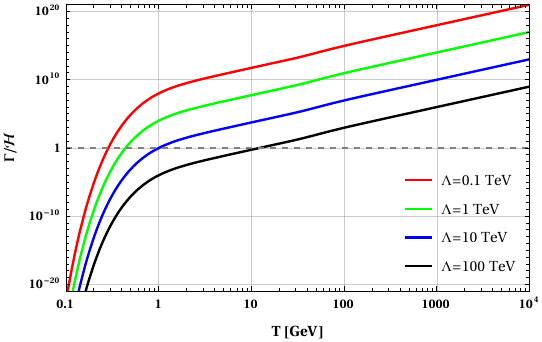}
	\includegraphics[width=0.45\linewidth]{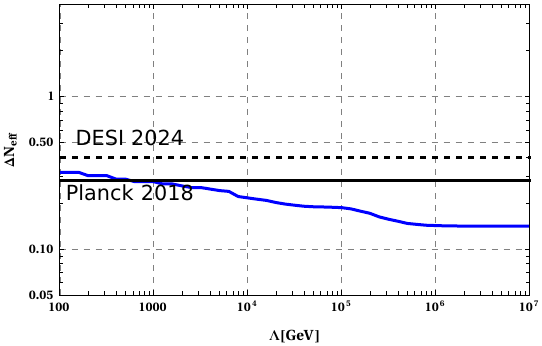}
	\caption{Left panel: comparison of $\nu_R$ interaction rate with Hubble rate of expansion assuming interactions of $\nu_{R}$ with SM quarks of 3rd generation only (case 5). Right panel: Variation of $\Delta N_{\rm eff}$ with cutoff scale $\Lambda$ for case 5.}
	\label{fig:Neff_3gen_quark}
\end{figure}

Here we consider the interaction of $\nu_{R}$ with the 3rd generation of quarks {\it i.e.} with top and bottom quarks, originated from $\mathcal{O}_{u\nu}$, $\mathcal{O}_{d\nu}$ and $\mathcal{O}_{Q\nu}$ operators. At a temperature below $m_{t}$, the $\nu_{R}$ interacts with $b$ quark only. After $T \lesssim m_{b}$, interaction rates get suppressed due to Boltzmann suppression of $b$ quark number density resulting a higher decoupling temperature. The contribution to $\Delta N_{\rm eff}$ for this scenario is shown in the right panel of Fig.~\ref{fig:Neff_3gen_quark}. Current constraint on $\Delta N_{\rm eff}$ from Planck 2018 (DESI 2024) exclude $\Lambda \lesssim 0.6$ TeV ($\Lambda \lesssim 0.001$) TeV.

\subsubsection{Interactions of $\nu_{R}$ with $\chi$ only}
\begin{figure}[hbt!]
	\centering
	\includegraphics[width=0.45\linewidth]{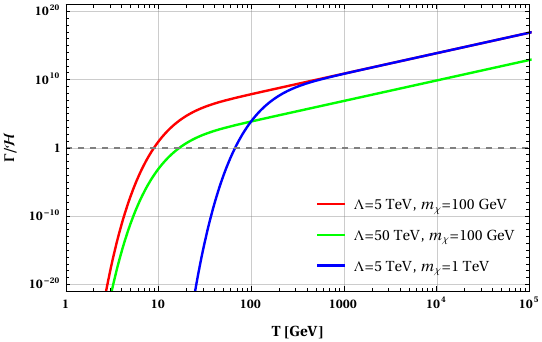}
	\includegraphics[width=0.45\linewidth]{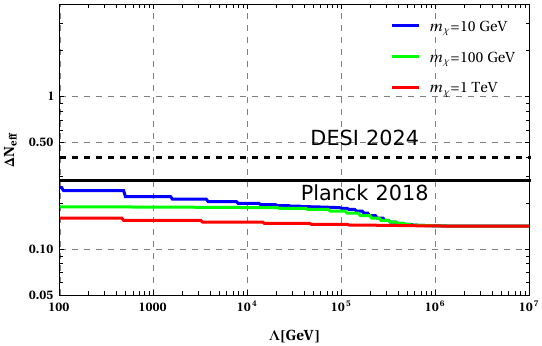}
	\caption{Left panel: comparison of $\nu_R$ interaction rate with Hubble rate of expansion assuming interactions of $\nu_{R}$ with DM only (case 6). Right panel: Variation of $\Delta N_{\rm eff}$ with cutoff scale $\Lambda$ for case 6.}
	\label{fig:Neff_DM}
\end{figure}

Here, we consider interaction of $\nu_{R}$ with $\chi$ via the operator $\mathcal{O}_{\chi\nu}$. Left panel of Fig. \ref{fig:Neff_DM} shows $\Gamma/\mathcal{H}$ variation with temperature for three sets of benchmark points ($\Lambda$ and $m_{\chi}$). Here, the decoupling of $\nu_{R}$ from the thermal bath is strongly dependent on $\chi$ mass due to Boltzmann suppression, whereas it is weakly dependent on the cut-off scale $\Lambda$. The right panel shows the $\Delta N_{\rm eff}$ values as a function of $\Lambda$ for three different $\chi$ masses.

\bibliographystyle{JHEP}
\bibliography{ref-eft, ref, ref1, ref1a, ref2, ref3}

\providecommand{\href}[2]{#2}\begingroup\raggedright\begin{thebibliography}{100}

\bibitem{Zyla:2020zbs}
{\scshape Particle Data Group} collaboration, \emph{{Review of Particle Physics}}, \href{https://doi.org/10.1093/ptep/ptaa104}{\emph{PTEP} {\bfseries 2020} (2020) 083C01}.

\bibitem{Planck:2018vyg}
{\scshape Planck} collaboration, \emph{{Planck 2018 results. VI. Cosmological parameters}}, \href{https://doi.org/10.1051/0004-6361/201833910}{\emph{Astron. Astrophys.} {\bfseries 641} (2020) A6} [\href{https://arxiv.org/abs/1807.06209}{{\ttfamily 1807.06209}}].

\bibitem{Roncadelli:1983ty}
M.~Roncadelli and D.~Wyler, \emph{{Naturally Light Dirac Neutrinos in Gauge Theories}}, \href{https://doi.org/10.1016/0370-2693(83)90156-9}{\emph{Phys. Lett. B} {\bfseries 133} (1983) 325}.

\bibitem{Roy:1983be}
P.~Roy and O.~U. Shanker, \emph{{Observable Neutrino Dirac Mass and Supergrand Unification}}, \href{https://doi.org/10.1103/PhysRevLett.52.713}{\emph{Phys. Rev. Lett.} {\bfseries 52} (1984) 713}.

\bibitem{Babu:1988yq}
K.~S. Babu and X.~G. He, \emph{{DIRAC NEUTRINO MASSES AS TWO LOOP RADIATIVE CORRECTIONS}}, \href{https://doi.org/10.1142/S0217732389000095}{\emph{Mod. Phys. Lett.} {\bfseries A4} (1989) 61}.

\bibitem{Peltoniemi:1992ss}
J.~T. Peltoniemi, D.~Tommasini and J.~W.~F. Valle, \emph{{Reconciling dark matter and solar neutrinos}}, \href{https://doi.org/10.1016/0370-2693(93)91837-D}{\emph{Phys. Lett.} {\bfseries B298} (1993) 383}.

\bibitem{Chulia:2016ngi}
S.~Centelles~Chuliá, E.~Ma, R.~Srivastava and J.~W.~F. Valle, \emph{{Dirac Neutrinos and Dark Matter Stability from Lepton Quarticity}}, \href{https://doi.org/10.1016/j.physletb.2017.01.070}{\emph{Phys. Lett.} {\bfseries B767} (2017) 209} [\href{https://arxiv.org/abs/1606.04543}{{\ttfamily 1606.04543}}].

\bibitem{Aranda:2013gga}
A.~Aranda, C.~Bonilla, S.~Morisi, E.~Peinado and J.~W.~F. Valle, \emph{{Dirac neutrinos from flavor symmetry}}, \href{https://doi.org/10.1103/PhysRevD.89.033001}{\emph{Phys. Rev.} {\bfseries D89} (2014) 033001} [\href{https://arxiv.org/abs/1307.3553}{{\ttfamily 1307.3553}}].

\bibitem{Chen:2015jta}
P.~Chen, G.-J. Ding, A.~D. Rojas, C.~A. Vaquera-Araujo and J.~W.~F. Valle, \emph{{Warped flavor symmetry predictions for neutrino physics}}, \href{https://doi.org/10.1007/JHEP01(2016)007}{\emph{JHEP} {\bfseries 01} (2016) 007} [\href{https://arxiv.org/abs/1509.06683}{{\ttfamily 1509.06683}}].

\bibitem{Ma:2015mjd}
E.~Ma, N.~Pollard, R.~Srivastava and M.~Zakeri, \emph{{Gauge $B-L$ Model with Residual $Z_3$ Symmetry}}, \href{https://doi.org/10.1016/j.physletb.2015.09.010}{\emph{Phys. Lett.} {\bfseries B750} (2015) 135} [\href{https://arxiv.org/abs/1507.03943}{{\ttfamily 1507.03943}}].

\bibitem{Reig:2016ewy}
M.~Reig, J.~W.~F. Valle and C.~A. Vaquera-Araujo, \emph{{Realistic $\mathrm{SU(3)_c \otimes SU(3)_L \otimes U(1)_X}$ model with a type II Dirac neutrino seesaw mechanism}}, \href{https://doi.org/10.1103/PhysRevD.94.033012}{\emph{Phys. Rev.} {\bfseries D94} (2016) 033012} [\href{https://arxiv.org/abs/1606.08499}{{\ttfamily 1606.08499}}].

\bibitem{Wang:2016lve}
W.~Wang and Z.-L. Han, \emph{{Naturally Small Dirac Neutrino Mass with Intermediate $SU(2)_{L}$ Multiplet Fields}},  \href{https://arxiv.org/abs/1611.03240}{{\ttfamily 1611.03240}}.

\bibitem{Wang:2017mcy}
W.~Wang, R.~Wang, Z.-L. Han and J.-Z. Han, \emph{{The $B-L$ Scotogenic Models for Dirac Neutrino Masses}}, \href{https://doi.org/10.1140/epjc/s10052-017-5446-9}{\emph{Eur. Phys. J.} {\bfseries C77} (2017) 889} [\href{https://arxiv.org/abs/1705.00414}{{\ttfamily 1705.00414}}].

\bibitem{Wang:2006jy}
F.~Wang, W.~Wang and J.~M. Yang, \emph{{Split two-Higgs-doublet model and neutrino condensation}}, \href{https://doi.org/10.1209/epl/i2006-10293-3}{\emph{Europhys. Lett.} {\bfseries 76} (2006) 388} [\href{https://arxiv.org/abs/hep-ph/0601018}{{\ttfamily hep-ph/0601018}}].

\bibitem{Gabriel:2006ns}
S.~Gabriel and S.~Nandi, \emph{{A New two Higgs doublet model}}, \href{https://doi.org/10.1016/j.physletb.2007.04.062}{\emph{Phys. Lett.} {\bfseries B655} (2007) 141} [\href{https://arxiv.org/abs/hep-ph/0610253}{{\ttfamily hep-ph/0610253}}].

\bibitem{Davidson:2009ha}
S.~M. Davidson and H.~E. Logan, \emph{{Dirac neutrinos from a second Higgs doublet}}, \href{https://doi.org/10.1103/PhysRevD.80.095008}{\emph{Phys. Rev.} {\bfseries D80} (2009) 095008} [\href{https://arxiv.org/abs/0906.3335}{{\ttfamily 0906.3335}}].

\bibitem{Davidson:2010sf}
S.~M. Davidson and H.~E. Logan, \emph{{LHC phenomenology of a two-Higgs-doublet neutrino mass model}}, \href{https://doi.org/10.1103/PhysRevD.82.115031}{\emph{Phys. Rev.} {\bfseries D82} (2010) 115031} [\href{https://arxiv.org/abs/1009.4413}{{\ttfamily 1009.4413}}].

\bibitem{Bonilla:2016zef}
C.~Bonilla and J.~W.~F. Valle, \emph{{Naturally light neutrinos in $Diracon$ model}}, \href{https://doi.org/10.1016/j.physletb.2016.09.022}{\emph{Phys. Lett.} {\bfseries B762} (2016) 162} [\href{https://arxiv.org/abs/1605.08362}{{\ttfamily 1605.08362}}].

\bibitem{Farzan:2012sa}
Y.~Farzan and E.~Ma, \emph{{Dirac neutrino mass generation from dark matter}}, \href{https://doi.org/10.1103/PhysRevD.86.033007}{\emph{Phys. Rev.} {\bfseries D86} (2012) 033007} [\href{https://arxiv.org/abs/1204.4890}{{\ttfamily 1204.4890}}].

\bibitem{Bonilla:2016diq}
C.~Bonilla, E.~Ma, E.~Peinado and J.~W.~F. Valle, \emph{{Two-loop Dirac neutrino mass and WIMP dark matter}}, \href{https://doi.org/10.1016/j.physletb.2016.09.027}{\emph{Phys. Lett.} {\bfseries B762} (2016) 214} [\href{https://arxiv.org/abs/1607.03931}{{\ttfamily 1607.03931}}].

\bibitem{Ma:2016mwh}
E.~Ma and O.~Popov, \emph{{Pathways to Naturally Small Dirac Neutrino Masses}}, \href{https://doi.org/10.1016/j.physletb.2016.11.027}{\emph{Phys. Lett.} {\bfseries B764} (2017) 142} [\href{https://arxiv.org/abs/1609.02538}{{\ttfamily 1609.02538}}].

\bibitem{Ma:2017kgb}
E.~Ma and U.~Sarkar, \emph{{Radiative Left-Right Dirac Neutrino Mass}}, \href{https://doi.org/10.1016/j.physletb.2017.08.071}{\emph{Phys. Lett.} {\bfseries B776} (2018) 54} [\href{https://arxiv.org/abs/1707.07698}{{\ttfamily 1707.07698}}].

\bibitem{Borah:2016lrl}
D.~Borah, \emph{{Light sterile neutrino and dark matter in left-right symmetric models without a Higgs bidoublet}}, \href{https://doi.org/10.1103/PhysRevD.94.075024}{\emph{Phys. Rev.} {\bfseries D94} (2016) 075024} [\href{https://arxiv.org/abs/1607.00244}{{\ttfamily 1607.00244}}].

\bibitem{Borah:2016zbd}
D.~Borah and A.~Dasgupta, \emph{{Common Origin of Neutrino Mass, Dark Matter and Dirac Leptogenesis}}, \href{https://doi.org/10.1088/1475-7516/2016/12/034}{\emph{JCAP} {\bfseries 1612} (2016) 034} [\href{https://arxiv.org/abs/1608.03872}{{\ttfamily 1608.03872}}].

\bibitem{Borah:2016hqn}
D.~Borah and A.~Dasgupta, \emph{{Observable Lepton Number Violation with Predominantly Dirac Nature of Active Neutrinos}}, \href{https://doi.org/10.1007/JHEP01(2017)072}{\emph{JHEP} {\bfseries 01} (2017) 072} [\href{https://arxiv.org/abs/1609.04236}{{\ttfamily 1609.04236}}].

\bibitem{Borah:2017leo}
D.~Borah and A.~Dasgupta, \emph{{Naturally Light Dirac Neutrino in Left-Right Symmetric Model}}, \href{https://doi.org/10.1088/1475-7516/2017/06/003}{\emph{JCAP} {\bfseries 1706} (2017) 003} [\href{https://arxiv.org/abs/1702.02877}{{\ttfamily 1702.02877}}].

\bibitem{CentellesChulia:2017koy}
S.~Centelles~Chuliá, R.~Srivastava and J.~W.~F. Valle, \emph{{Generalized Bottom-Tau unification, neutrino oscillations and dark matter: predictions from a lepton quarticity flavor approach}}, \href{https://doi.org/10.1016/j.physletb.2017.07.065}{\emph{Phys. Lett.} {\bfseries B773} (2017) 26} [\href{https://arxiv.org/abs/1706.00210}{{\ttfamily 1706.00210}}].

\bibitem{Bonilla:2017ekt}
C.~Bonilla, J.~M. Lamprea, E.~Peinado and J.~W.~F. Valle, \emph{{Flavour-symmetric type-II Dirac neutrino seesaw mechanism}}, \href{https://doi.org/10.1016/j.physletb.2018.02.022}{\emph{Phys. Lett.} {\bfseries B779} (2018) 257} [\href{https://arxiv.org/abs/1710.06498}{{\ttfamily 1710.06498}}].

\bibitem{Memenga:2013vc}
N.~Memenga, W.~Rodejohann and H.~Zhang, \emph{{$A_4$ flavor symmetry model for Dirac neutrinos and sizable $U_{e3}$}}, \href{https://doi.org/10.1103/PhysRevD.87.053021}{\emph{Phys. Rev.} {\bfseries D87} (2013) 053021} [\href{https://arxiv.org/abs/1301.2963}{{\ttfamily 1301.2963}}].

\bibitem{Borah:2017dmk}
D.~Borah and B.~Karmakar, \emph{{$A_4$ flavour model for Dirac neutrinos: Type I and inverse seesaw}}, \href{https://doi.org/10.1016/j.physletb.2018.03.047}{\emph{Phys. Lett.} {\bfseries B780} (2018) 461} [\href{https://arxiv.org/abs/1712.06407}{{\ttfamily 1712.06407}}].

\bibitem{CentellesChulia:2018gwr}
S.~Centelles~Chuliá, R.~Srivastava and J.~W.~F. Valle, \emph{{Seesaw roadmap to neutrino mass and dark matter}}, \href{https://doi.org/10.1016/j.physletb.2018.03.046}{\emph{Phys. Lett.} {\bfseries B781} (2018) 122} [\href{https://arxiv.org/abs/1802.05722}{{\ttfamily 1802.05722}}].

\bibitem{CentellesChulia:2018bkz}
S.~Centelles~Chuliá, R.~Srivastava and J.~W.~F. Valle, \emph{{Seesaw Dirac neutrino mass through dimension-6 operators}},  \href{https://arxiv.org/abs/1804.03181}{{\ttfamily 1804.03181}}.

\bibitem{Han:2018zcn}
Z.-L. Han and W.~Wang, \emph{{$Z'$ Portal Dark Matter in $B-L$ Scotogenic Dirac Model}},  \href{https://arxiv.org/abs/1805.02025}{{\ttfamily 1805.02025}}.

\bibitem{Borah:2018gjk}
D.~Borah, B.~Karmakar and D.~Nanda, \emph{{Common Origin of Dirac Neutrino Mass and Freeze-in Massive Particle Dark Matter}}, \href{https://doi.org/10.1088/1475-7516/2018/07/039}{\emph{JCAP} {\bfseries 1807} (2018) 039} [\href{https://arxiv.org/abs/1805.11115}{{\ttfamily 1805.11115}}].

\bibitem{Borah:2018nvu}
D.~Borah and B.~Karmakar, \emph{{Linear seesaw for Dirac neutrinos with $A_4$ flavour symmetry}}, \href{https://doi.org/10.1016/j.physletb.2018.12.006}{\emph{Phys. Lett.} {\bfseries B789} (2019) 59} [\href{https://arxiv.org/abs/1806.10685}{{\ttfamily 1806.10685}}].

\bibitem{Borah:2019bdi}
D.~Borah, D.~Nanda and A.~K. Saha, \emph{{Common origin of modified chaotic inflation, non thermal dark matter and Dirac neutrino mass}},  \href{https://arxiv.org/abs/1904.04840}{{\ttfamily 1904.04840}}.

\bibitem{CentellesChulia:2019xky}
S.~Centelles~Chuliá, R.~Cepedello, E.~Peinado and R.~Srivastava, \emph{{Systematic classification of two loop $d$ = 4 Dirac neutrino mass models and the Diracness-dark matter stability connection}}, \href{https://doi.org/10.1007/JHEP10(2019)093}{\emph{JHEP} {\bfseries 10} (2019) 093} [\href{https://arxiv.org/abs/1907.08630}{{\ttfamily 1907.08630}}].

\bibitem{Jana:2019mgj}
S.~Jana, V.~P. K. and S.~Saad, \emph{{Minimal Realizations of Dirac Neutrino Mass from Generic One-loop and Two-loop Topologies at $d=5$}},  \href{https://arxiv.org/abs/1910.09537}{{\ttfamily 1910.09537}}.

\bibitem{Nanda:2019nqy}
D.~Nanda and D.~Borah, \emph{{Connecting Light Dirac Neutrinos to a Multi-component Dark Matter Scenario in Gauged $B-L$ Model}},  \href{https://arxiv.org/abs/1911.04703}{{\ttfamily 1911.04703}}.

\bibitem{Guo:2020qin}
S.-Y. Guo and Z.-L. Han, \emph{{Observable Signatures of Scotogenic Dirac Model}}, \href{https://doi.org/10.1007/JHEP12(2020)062}{\emph{JHEP} {\bfseries 12} (2020) 062} [\href{https://arxiv.org/abs/2005.08287}{{\ttfamily 2005.08287}}].

\bibitem{Bernal:2021ezl}
N.~Bernal, J.~Calle and D.~Restrepo, \emph{{Anomaly-free Abelian gauge symmetries with Dirac scotogenic models}}, \href{https://doi.org/10.1103/PhysRevD.103.095032}{\emph{Phys. Rev. D} {\bfseries 103} (2021) 095032} [\href{https://arxiv.org/abs/2102.06211}{{\ttfamily 2102.06211}}].

\bibitem{Borah:2022obi}
D.~Borah, S.~Mahapatra, D.~Nanda and N.~Sahu, \emph{{Type II Dirac Seesaw with Observable $\Delta N_{\rm eff}$ in the light of W-mass Anomaly}},  \href{https://arxiv.org/abs/2204.08266}{{\ttfamily 2204.08266}}.

\bibitem{Li:2022chc}
S.-P. Li, X.-Q. Li, X.-S. Yan and Y.-D. Yang, \emph{{Scotogenic Dirac neutrino mass models embedded with leptoquarks: one pathway to address the flavor anomalies and the neutrino masses together}}, \href{https://doi.org/10.1140/epjc/s10052-022-11054-w}{\emph{Eur. Phys. J. C} {\bfseries 82} (2022) 1078} [\href{https://arxiv.org/abs/2204.09201}{{\ttfamily 2204.09201}}].

\bibitem{Dey:2024ctx}
M.~Dey and S.~Roy, \emph{{Revisiting the Dirac Nature of Neutrinos}},  \href{https://arxiv.org/abs/2403.12461}{{\ttfamily 2403.12461}}.

\bibitem{Singh:2024imk}
L.~Singh, M.~Kashav and S.~Verma, \emph{{Minimal Type-I Dirac seesaw and Leptogenesis under $A_{4}$ modular invariance}},  \href{https://arxiv.org/abs/2405.07165}{{\ttfamily 2405.07165}}.

\bibitem{Borah:2024gql}
D.~Borah, P.~Das, B.~Karmakar and S.~Mahapatra, \emph{{Discrete dark matter with light Dirac neutrinos}},  \href{https://arxiv.org/abs/2406.17861}{{\ttfamily 2406.17861}}.

\bibitem{Beltran:2008xg}
M.~Beltran, D.~Hooper, E.~W. Kolb and Z.~C. Krusberg, \emph{{Deducing the nature of dark matter from direct and indirect detection experiments in the absence of collider signatures of new physics}}, \href{https://doi.org/10.1103/PhysRevD.80.043509}{\emph{Phys. Rev. D} {\bfseries 80} (2009) 043509} [\href{https://arxiv.org/abs/0808.3384}{{\ttfamily 0808.3384}}].

\bibitem{Fan:2010gt}
J.~Fan, M.~Reece and L.-T. Wang, \emph{{Non-relativistic effective theory of dark matter direct detection}}, \href{https://doi.org/10.1088/1475-7516/2010/11/042}{\emph{JCAP} {\bfseries 11} (2010) 042} [\href{https://arxiv.org/abs/1008.1591}{{\ttfamily 1008.1591}}].

\bibitem{Goodman:2010ku}
J.~Goodman, M.~Ibe, A.~Rajaraman, W.~Shepherd, T.~M. Tait and H.-B. Yu, \emph{{Constraints on Dark Matter from Colliders}}, \href{https://doi.org/10.1103/PhysRevD.82.116010}{\emph{Phys. Rev. D} {\bfseries 82} (2010) 116010} [\href{https://arxiv.org/abs/1008.1783}{{\ttfamily 1008.1783}}].

\bibitem{Beltran:2010ww}
M.~Beltran, D.~Hooper, E.~W. Kolb, Z.~A.~C. Krusberg and T.~M.~P. Tait, \emph{{Maverick dark matter at colliders}}, \href{https://doi.org/10.1007/JHEP09(2010)037}{\emph{JHEP} {\bfseries 09} (2010) 037} [\href{https://arxiv.org/abs/1002.4137}{{\ttfamily 1002.4137}}].

\bibitem{Fitzpatrick:2012ix}
A.~L. Fitzpatrick, W.~Haxton, E.~Katz, N.~Lubbers and Y.~Xu, \emph{{The Effective Field Theory of Dark Matter Direct Detection}}, \href{https://doi.org/10.1088/1475-7516/2013/02/004}{\emph{JCAP} {\bfseries 02} (2013) 004} [\href{https://arxiv.org/abs/1203.3542}{{\ttfamily 1203.3542}}].

\bibitem{Bhattacharya:2021edh}
S.~Bhattacharya and J.~Wudka, \emph{{Effective theories with dark matter applications}}, \href{https://doi.org/10.1142/S0218271821300044}{\emph{Int. J. Mod. Phys. D} {\bfseries 30} (2021) 2130004} [\href{https://arxiv.org/abs/2104.01788}{{\ttfamily 2104.01788}}].

\bibitem{Luo:2020sho}
X.~Luo, W.~Rodejohann and X.-J. Xu, \emph{{Dirac neutrinos and $N_{{\rm eff}}$}}, \href{https://doi.org/10.1088/1475-7516/2020/06/058}{\emph{JCAP} {\bfseries 06} (2020) 058} [\href{https://arxiv.org/abs/2005.01629}{{\ttfamily 2005.01629}}].

\bibitem{Luo:2020fdt}
X.~Luo, W.~Rodejohann and X.-J. Xu, \emph{{Dirac neutrinos and $N_{{\rm eff}}$ II: the freeze-in case}},  \href{https://arxiv.org/abs/2011.13059}{{\ttfamily 2011.13059}}.

\bibitem{Brust:2013ova}
C.~Brust, D.~E. Kaplan and M.~T. Walters, \emph{{New Light Species and the CMB}}, \href{https://doi.org/10.1007/JHEP12(2013)058}{\emph{JHEP} {\bfseries 12} (2013) 058} [\href{https://arxiv.org/abs/1303.5379}{{\ttfamily 1303.5379}}].

\bibitem{Du:2021idh}
Y.~Du and J.-H. Yu, \emph{{Neutrino non-standard interactions meet precision measurements of N$_{eff}$}}, \href{https://doi.org/10.1007/JHEP05(2021)058}{\emph{JHEP} {\bfseries 05} (2021) 058} [\href{https://arxiv.org/abs/2101.10475}{{\ttfamily 2101.10475}}].

\bibitem{LZ:2022lsv}
{\scshape LZ} collaboration, \emph{{First Dark Matter Search Results from the LUX-ZEPLIN (LZ) Experiment}}, \href{https://doi.org/10.1103/PhysRevLett.131.041002}{\emph{Phys. Rev. Lett.} {\bfseries 131} (2023) 041002} [\href{https://arxiv.org/abs/2207.03764}{{\ttfamily 2207.03764}}].

\bibitem{LZ:2024}
{\scshape LZ} collaboration, \emph{{New Dark Matter Search Results from the LUX-ZEPLIN (LZ) Experiment}}, \href{https://doi.org/https://indico.uchicago.edu/event/427/contributions/1325/}{\emph{Plenary Talk at TeVPA 2024} }.

\bibitem{Fox:2011fx}
P.~J. Fox, R.~Harnik, J.~Kopp and Y.~Tsai, \emph{{LEP Shines Light on Dark Matter}}, \href{https://doi.org/10.1103/PhysRevD.84.014028}{\emph{Phys. Rev. D} {\bfseries 84} (2011) 014028} [\href{https://arxiv.org/abs/1103.0240}{{\ttfamily 1103.0240}}].

\bibitem{Essig:2013vha}
R.~Essig, J.~Mardon, M.~Papucci, T.~Volansky and Y.-M. Zhong, \emph{{Constraining Light Dark Matter with Low-Energy $e^+e^-$ Colliders}}, \href{https://doi.org/10.1007/JHEP11(2013)167}{\emph{JHEP} {\bfseries 11} (2013) 167} [\href{https://arxiv.org/abs/1309.5084}{{\ttfamily 1309.5084}}].

\bibitem{Yu:2013aca}
Z.-H. Yu, Q.-S. Yan and P.-F. Yin, \emph{{Detecting interactions between dark matter and photons at high energy $e^+e^-$ colliders}}, \href{https://doi.org/10.1103/PhysRevD.88.075015}{\emph{Phys. Rev. D} {\bfseries 88} (2013) 075015} [\href{https://arxiv.org/abs/1307.5740}{{\ttfamily 1307.5740}}].

\bibitem{Kadota:2014mea}
K.~Kadota and J.~Silk, \emph{{Constraints on Light Magnetic Dipole Dark Matter from the ILC and SN 1987A}}, \href{https://doi.org/10.1103/PhysRevD.89.103528}{\emph{Phys. Rev. D} {\bfseries 89} (2014) 103528} [\href{https://arxiv.org/abs/1402.7295}{{\ttfamily 1402.7295}}].

\bibitem{Yu:2014ula}
Z.-H. Yu, X.-J. Bi, Q.-S. Yan and P.-F. Yin, \emph{{Dark matter searches in the mono-$Z$ channel at high energy $e^+e^-$ colliders}}, \href{https://doi.org/10.1103/PhysRevD.90.055010}{\emph{Phys. Rev. D} {\bfseries 90} (2014) 055010} [\href{https://arxiv.org/abs/1404.6990}{{\ttfamily 1404.6990}}].

\bibitem{Freitas:2014jla}
A.~Freitas and S.~Westhoff, \emph{{Leptophilic Dark Matter in Lepton Interactions at LEP and ILC}}, \href{https://doi.org/10.1007/JHEP10(2014)116}{\emph{JHEP} {\bfseries 10} (2014) 116} [\href{https://arxiv.org/abs/1408.1959}{{\ttfamily 1408.1959}}].

\bibitem{Dutta:2017ljq}
S.~Dutta, D.~Sachdeva and B.~Rawat, \emph{{Signals of Leptophilic Dark Matter at the ILC}}, \href{https://doi.org/10.1140/epjc/s10052-017-5188-8}{\emph{Eur. Phys. J. C} {\bfseries 77} (2017) 639} [\href{https://arxiv.org/abs/1704.03994}{{\ttfamily 1704.03994}}].

\bibitem{Liu:2019ogn}
Z.~Liu, Y.-H. Xu and Y.~Zhang, \emph{{Probing dark matter particles at CEPC}}, \href{https://doi.org/10.1007/JHEP06(2019)009}{\emph{JHEP} {\bfseries 06} (2019) 009} [\href{https://arxiv.org/abs/1903.12114}{{\ttfamily 1903.12114}}].

\bibitem{Choudhury:2019sxt}
D.~Choudhury and D.~Sachdeva, \emph{{Model independent analysis of MeV scale dark matter. II. Implications from $e^-e^+$ colliders and direct detection}}, \href{https://doi.org/10.1103/PhysRevD.100.075007}{\emph{Phys. Rev. D} {\bfseries 100} (2019) 075007} [\href{https://arxiv.org/abs/1906.06364}{{\ttfamily 1906.06364}}].

\bibitem{Kundu:2021cmo}
S.~Kundu, A.~Guha, P.~K. Das and P.~S.~B. Dev, \emph{{EFT analysis of leptophilic dark matter at future electron-positron colliders in the mono-photon and mono-Z channels}}, \href{https://doi.org/10.1103/PhysRevD.107.015003}{\emph{Phys. Rev. D} {\bfseries 107} (2023) 015003} [\href{https://arxiv.org/abs/2110.06903}{{\ttfamily 2110.06903}}].

\bibitem{Barman:2021hhg}
B.~Barman, S.~Bhattacharya, S.~Girmohanta and S.~Jahedi, \emph{{Effective Leptophilic WIMPs at the $e^{+}e^{-}$ collider}}, \href{https://doi.org/10.1007/JHEP04(2022)146}{\emph{JHEP} {\bfseries 04} (2022) 146} [\href{https://arxiv.org/abs/2109.10936}{{\ttfamily 2109.10936}}].

\bibitem{Bhattacharya:2022wtr}
S.~Bhattacharya, P.~Ghosh, J.~Lahiri and B.~Mukhopadhyaya, \emph{{Distinguishing two dark matter component particles at $e^{+}e^{-}$ colliders}}, \href{https://doi.org/10.1007/JHEP12(2022)049}{\emph{JHEP} {\bfseries 12} (2022) 049} [\href{https://arxiv.org/abs/2202.12097}{{\ttfamily 2202.12097}}].

\bibitem{Bhattacharya:2022qck}
S.~Bhattacharya, P.~Ghosh, J.~Lahiri and B.~Mukhopadhyaya, \emph{{Mono-X signal and two component dark matter: New distinction criteria}}, \href{https://doi.org/10.1103/PhysRevD.108.L111703}{\emph{Phys. Rev. D} {\bfseries 108} (2023) L111703} [\href{https://arxiv.org/abs/2211.10749}{{\ttfamily 2211.10749}}].

\bibitem{Ge:2023wye}
S.-F. Ge, K.~Ma, X.-D. Ma and J.~Sheng, \emph{{Associated production of neutrino and dark fermion at future lepton colliders}}, \href{https://doi.org/10.1007/JHEP11(2023)190}{\emph{JHEP} {\bfseries 11} (2023) 190} [\href{https://arxiv.org/abs/2306.00657}{{\ttfamily 2306.00657}}].

\bibitem{Roy:2024ear}
A.~Roy, B.~Dasgupta and M.~Guchait, \emph{{Constraining Asymmetric Dark Matter using colliders and direct detection}}, \href{https://doi.org/10.1007/JHEP08(2024)095}{\emph{JHEP} {\bfseries 08} (2024) 095} [\href{https://arxiv.org/abs/2402.17265}{{\ttfamily 2402.17265}}].

\bibitem{Barman:2024nhr}
B.~Barman, S.~Bhattacharya, S.~Jahedi, D.~Pradhan and A.~Sarkar, \emph{{Lepton Collider as a window to Reheating}},  \href{https://arxiv.org/abs/2406.11963}{{\ttfamily 2406.11963}}.

\bibitem{Han:2020uak}
T.~Han, Z.~Liu, L.-T. Wang and X.~Wang, \emph{{WIMPs at High Energy Muon Colliders}}, \href{https://doi.org/10.1103/PhysRevD.103.075004}{\emph{Phys. Rev. D} {\bfseries 103} (2021) 075004} [\href{https://arxiv.org/abs/2009.11287}{{\ttfamily 2009.11287}}].

\bibitem{Abazajian:2019eic}
K.~Abazajian et~al., \emph{{CMB-S4 Science Case, Reference Design, and Project Plan}},  \href{https://arxiv.org/abs/1907.04473}{{\ttfamily 1907.04473}}.

\bibitem{CMB-HD:2022bsz}
{\scshape CMB-HD} collaboration, \emph{{Snowmass2021 CMB-HD White Paper}},  \href{https://arxiv.org/abs/2203.05728}{{\ttfamily 2203.05728}}.

\bibitem{Bhattacharya:2015vja}
S.~Bhattacharya and J.~Wudka, \emph{{Dimension-seven operators in the standard model with right handed neutrinos}}, \href{https://doi.org/10.1103/PhysRevD.94.055022}{\emph{Phys. Rev. D} {\bfseries 94} (2016) 055022} [\href{https://arxiv.org/abs/1505.05264}{{\ttfamily 1505.05264}}].

\bibitem{Liao:2016qyd}
Y.~Liao and X.-D. Ma, \emph{{Operators up to Dimension Seven in Standard Model Effective Field Theory Extended with Sterile Neutrinos}}, \href{https://doi.org/10.1103/PhysRevD.96.015012}{\emph{Phys. Rev. D} {\bfseries 96} (2017) 015012} [\href{https://arxiv.org/abs/1612.04527}{{\ttfamily 1612.04527}}].

\bibitem{Kolb:1990vq}
E.~W. Kolb and M.~S. Turner, \emph{{The Early Universe}},  1990.
\newblock 10.1201/9780429492860.

\bibitem{Gondolo:1990dk}
P.~Gondolo and G.~Gelmini, \emph{{Cosmic abundances of stable particles: Improved analysis}}, \href{https://doi.org/10.1016/0550-3213(91)90438-4}{\emph{Nucl. Phys.} {\bfseries B360} (1991) 145}.

\bibitem{Guo:2010hq}
W.-L. Guo and Y.-L. Wu, \emph{{The Real singlet scalar dark matter model}}, \href{https://doi.org/10.1007/JHEP10(2010)083}{\emph{JHEP} {\bfseries 10} (2010) 083} [\href{https://arxiv.org/abs/1006.2518}{{\ttfamily 1006.2518}}].

\bibitem{Kopp:2009et}
J.~Kopp, V.~Niro, T.~Schwetz and J.~Zupan, \emph{{DAMA/LIBRA and leptonically interacting Dark Matter}}, \href{https://doi.org/10.1103/PhysRevD.80.083502}{\emph{Phys. Rev. D} {\bfseries 80} (2009) 083502} [\href{https://arxiv.org/abs/0907.3159}{{\ttfamily 0907.3159}}].

\bibitem{Fermi-LAT:2016afa}
{\scshape Fermi-LAT} collaboration, \emph{{Sensitivity Projections for Dark Matter Searches with the Fermi Large Area Telescope}}, \href{https://doi.org/10.1016/j.physrep.2016.05.001}{\emph{Phys. Rept.} {\bfseries 636} (2016) 1} [\href{https://arxiv.org/abs/1605.02016}{{\ttfamily 1605.02016}}].

\bibitem{CTAConsortium:2012fwj}
{\scshape CTA Consortium} collaboration, \emph{{Dark Matter and Fundamental Physics with the Cherenkov Telescope Array}}, \href{https://doi.org/10.1016/j.astropartphys.2012.08.002}{\emph{Astropart. Phys.} {\bfseries 43} (2013) 189} [\href{https://arxiv.org/abs/1208.5356}{{\ttfamily 1208.5356}}].

\bibitem{Behnke:2013xla}
\emph{{The International Linear Collider Technical Design Report - Volume 1: Executive Summary}},  \href{https://arxiv.org/abs/1306.6327}{{\ttfamily 1306.6327}}.

\bibitem{Black:2022cth}
K.~M. Black et~al., \emph{{Muon Collider Forum Report}},  \href{https://arxiv.org/abs/2209.01318}{{\ttfamily 2209.01318}}.

\bibitem{Alwall:2014hca}
J.~Alwall, R.~Frederix, S.~Frixione, V.~Hirschi, F.~Maltoni, O.~Mattelaer et~al., \emph{{The automated computation of tree-level and next-to-leading order differential cross sections, and their matching to parton shower simulations}}, \href{https://doi.org/10.1007/JHEP07(2014)079}{\emph{JHEP} {\bfseries 07} (2014) 079} [\href{https://arxiv.org/abs/1405.0301}{{\ttfamily 1405.0301}}].

\bibitem{Sjostrand:2014zea}
T.~Sj\"ostrand, S.~Ask, J.~R. Christiansen, R.~Corke, N.~Desai, P.~Ilten et~al., \emph{{An introduction to PYTHIA 8.2}}, \href{https://doi.org/10.1016/j.cpc.2015.01.024}{\emph{Comput. Phys. Commun.} {\bfseries 191} (2015) 159} [\href{https://arxiv.org/abs/1410.3012}{{\ttfamily 1410.3012}}].

\bibitem{deFavereau:2013fsa}
{\scshape DELPHES 3} collaboration, \emph{{DELPHES 3, A modular framework for fast simulation of a generic collider experiment}}, \href{https://doi.org/10.1007/JHEP02(2014)057}{\emph{JHEP} {\bfseries 02} (2014) 057} [\href{https://arxiv.org/abs/1307.6346}{{\ttfamily 1307.6346}}].

\bibitem{Christensen:2008py}
N.~D. Christensen and C.~Duhr, \emph{{FeynRules - Feynman rules made easy}}, \href{https://doi.org/10.1016/j.cpc.2009.02.018}{\emph{Comput. Phys. Commun.} {\bfseries 180} (2009) 1614} [\href{https://arxiv.org/abs/0806.4194}{{\ttfamily 0806.4194}}].

\bibitem{Agapov:2022bhm}
I.~Agapov et~al., \emph{{Future Circular Lepton Collider FCC-ee: Overview and Status}},  in \emph{{Snowmass 2021}}, 3, 2022, \href{https://arxiv.org/abs/2203.08310}{{\ttfamily 2203.08310}}.

\bibitem{Brunner:2022usy}
O.~Brunner et~al., \emph{{The CLIC project}},  \href{https://arxiv.org/abs/2203.09186}{{\ttfamily 2203.09186}}.

\bibitem{Assmann:2002th}
R.~Assmann, M.~Lamont and S.~Myers, \emph{{A brief history of the LEP collider}}, \href{https://doi.org/10.1016/S0920-5632(02)90005-8}{\emph{Nucl. Phys. B Proc. Suppl.} {\bfseries 109} (2002) 17}.

\bibitem{Abazajian:2019oqj}
K.~N. Abazajian and J.~Heeck, \emph{{Observing Dirac neutrinos in the cosmic microwave background}}, \href{https://doi.org/10.1103/PhysRevD.100.075027}{\emph{Phys. Rev.} {\bfseries D100} (2019) 075027} [\href{https://arxiv.org/abs/1908.03286}{{\ttfamily 1908.03286}}].

\bibitem{FileviezPerez:2019cyn}
P.~Fileviez~Pérez, C.~Murgui and A.~D. Plascencia, \emph{{Neutrino-Dark Matter Connections in Gauge Theories}}, \href{https://doi.org/10.1103/PhysRevD.100.035041}{\emph{Phys. Rev.} {\bfseries D100} (2019) 035041} [\href{https://arxiv.org/abs/1905.06344}{{\ttfamily 1905.06344}}].

\bibitem{Han:2020oet}
C.~Han, M.~López-Ibáñez, B.~Peng and J.~M. Yang, \emph{{Dirac dark matter in $U(1)_{B-L}$ with Stueckelberg mechanism}},  \href{https://arxiv.org/abs/2001.04078}{{\ttfamily 2001.04078}}.

\bibitem{Borah:2020boy}
D.~Borah, A.~Dasgupta, C.~Majumdar and D.~Nanda, \emph{{Observing left-right symmetry in the cosmic microwave background}}, \href{https://doi.org/10.1103/PhysRevD.102.035025}{\emph{Phys. Rev. D} {\bfseries 102} (2020) 035025} [\href{https://arxiv.org/abs/2005.02343}{{\ttfamily 2005.02343}}].

\bibitem{Adshead:2020ekg}
P.~Adshead, Y.~Cui, A.~J. Long and M.~Shamma, \emph{{Unraveling the Dirac Neutrino with Cosmological and Terrestrial Detectors}},  \href{https://arxiv.org/abs/2009.07852}{{\ttfamily 2009.07852}}.

\bibitem{Mahanta:2021plx}
D.~Mahanta and D.~Borah, \emph{{Low scale Dirac leptogenesis and dark matter with observable $\Delta N_{\rm eff}$}},  \href{https://arxiv.org/abs/2101.02092}{{\ttfamily 2101.02092}}.

\bibitem{Biswas:2021kio}
A.~Biswas, D.~Borah and D.~Nanda, \emph{{Light Dirac neutrino portal dark matter with observable \ensuremath{\Delta}Neff}}, \href{https://doi.org/10.1088/1475-7516/2021/10/002}{\emph{JCAP} {\bfseries 10} (2021) 002} [\href{https://arxiv.org/abs/2103.05648}{{\ttfamily 2103.05648}}].

\bibitem{Borah:2022qln}
D.~Borah, S.~Jyoti~Das and N.~Okada, \emph{{Affleck-Dine cogenesis of baryon and dark matter}}, \href{https://doi.org/10.1007/JHEP05(2023)004}{\emph{JHEP} {\bfseries 05} (2023) 004} [\href{https://arxiv.org/abs/2212.04516}{{\ttfamily 2212.04516}}].

\bibitem{Li:2022yna}
S.-P. Li, X.-Q. Li, X.-S. Yan and Y.-D. Yang, \emph{{Effective neutrino number shift from keV-vacuum neutrinophilic 2HDM}},  \href{https://arxiv.org/abs/2202.10250}{{\ttfamily 2202.10250}}.

\bibitem{Biswas:2022fga}
A.~Biswas, D.~K. Ghosh and D.~Nanda, \emph{{Concealing Dirac neutrinos from cosmic microwave background}}, \href{https://doi.org/10.1088/1475-7516/2022/10/006}{\emph{JCAP} {\bfseries 10} (2022) 006} [\href{https://arxiv.org/abs/2206.13710}{{\ttfamily 2206.13710}}].

\bibitem{Adshead:2022ovo}
P.~Adshead, P.~Ralegankar and J.~Shelton, \emph{{Dark radiation constraints on portal interactions with hidden sectors}}, \href{https://doi.org/10.1088/1475-7516/2022/09/056}{\emph{JCAP} {\bfseries 09} (2022) 056} [\href{https://arxiv.org/abs/2206.13530}{{\ttfamily 2206.13530}}].

\bibitem{Borah:2023dhk}
D.~Borah, S.~Mahapatra, D.~Nanda, S.~K. Sahoo and N.~Sahu, \emph{{Singlet-doublet fermion dark matter with Dirac neutrino mass, $(g-2)_\mu$ and $\Delta N_{\rm eff}$}},  \href{https://arxiv.org/abs/2310.03721}{{\ttfamily 2310.03721}}.

\bibitem{Borah:2022enh}
D.~Borah, P.~Das and D.~Nanda, \emph{{Observable ${\rm \Delta{N_{eff}}}$ in Dirac Scotogenic Model}},  \href{https://arxiv.org/abs/2211.13168}{{\ttfamily 2211.13168}}.

\bibitem{Das:2023oph}
N.~Das, S.~Jyoti~Das and D.~Borah, \emph{{Thermalized dark radiation in the presence of a PBH: \ensuremath{\Delta}Neff and gravitational waves complementarity}}, \href{https://doi.org/10.1103/PhysRevD.108.095052}{\emph{Phys. Rev. D} {\bfseries 108} (2023) 095052} [\href{https://arxiv.org/abs/2306.00067}{{\ttfamily 2306.00067}}].

\bibitem{Esseili:2023ldf}
H.~Esseili and G.~D. Kribs, \emph{{Cosmological implications of gauged U(1)$_{B-L}$ on \ensuremath{\Delta}N $_{eff}$ in the CMB and BBN}}, \href{https://doi.org/10.1088/1475-7516/2024/05/110}{\emph{JCAP} {\bfseries 05} (2024) 110} [\href{https://arxiv.org/abs/2308.07955}{{\ttfamily 2308.07955}}].

\bibitem{Das:2023yhv}
N.~Das and D.~Borah, \emph{{Light Dirac neutrino portal dark matter with gauged U(1)B-L symmetry}}, \href{https://doi.org/10.1103/PhysRevD.109.075045}{\emph{Phys. Rev. D} {\bfseries 109} (2024) 075045} [\href{https://arxiv.org/abs/2312.06777}{{\ttfamily 2312.06777}}].

\bibitem{Mangano:2005cc}
G.~Mangano, G.~Miele, S.~Pastor, T.~Pinto, O.~Pisanti and P.~D. Serpico, \emph{{Relic neutrino decoupling including flavor oscillations}}, \href{https://doi.org/10.1016/j.nuclphysb.2005.09.041}{\emph{Nucl. Phys. B} {\bfseries 729} (2005) 221} [\href{https://arxiv.org/abs/hep-ph/0506164}{{\ttfamily hep-ph/0506164}}].

\bibitem{Grohs:2015tfy}
E.~Grohs, G.~M. Fuller, C.~T. Kishimoto, M.~W. Paris and A.~Vlasenko, \emph{{Neutrino energy transport in weak decoupling and big bang nucleosynthesis}}, \href{https://doi.org/10.1103/PhysRevD.93.083522}{\emph{Phys. Rev. D} {\bfseries 93} (2016) 083522} [\href{https://arxiv.org/abs/1512.02205}{{\ttfamily 1512.02205}}].

\bibitem{deSalas:2016ztq}
P.~F. de~Salas and S.~Pastor, \emph{{Relic neutrino decoupling with flavour oscillations revisited}}, \href{https://doi.org/10.1088/1475-7516/2016/07/051}{\emph{JCAP} {\bfseries 1607} (2016) 051} [\href{https://arxiv.org/abs/1606.06986}{{\ttfamily 1606.06986}}].

\bibitem{DESI:2024mwx}
{\scshape DESI} collaboration, \emph{{DESI 2024 VI: Cosmological Constraints from the Measurements of Baryon Acoustic Oscillations}},  \href{https://arxiv.org/abs/2404.03002}{{\ttfamily 2404.03002}}.

\bibitem{Cyburt:2015mya}
R.~H. Cyburt, B.~D. Fields, K.~A. Olive and T.-H. Yeh, \emph{{Big Bang Nucleosynthesis: 2015}}, \href{https://doi.org/10.1103/RevModPhys.88.015004}{\emph{Rev. Mod. Phys.} {\bfseries 88} (2016) 015004} [\href{https://arxiv.org/abs/1505.01076}{{\ttfamily 1505.01076}}].

\bibitem{He:1990pn}
X.~G. He, G.~C. Joshi, H.~Lew and R.~R. Volkas, \emph{{NEW Z-prime PHENOMENOLOGY}}, \href{https://doi.org/10.1103/PhysRevD.43.R22}{\emph{Phys. Rev. D} {\bfseries 43} (1991) 22}.

\bibitem{Ma:2001md}
E.~Ma, D.~P. Roy and S.~Roy, \emph{{Gauged L(mu) - L(tau) with large muon anomalous magnetic moment and the bimaximal mixing of neutrinos}}, \href{https://doi.org/10.1016/S0370-2693(01)01428-9}{\emph{Phys. Lett. B} {\bfseries 525} (2002) 101} [\href{https://arxiv.org/abs/hep-ph/0110146}{{\ttfamily hep-ph/0110146}}].

\bibitem{Baek:2008nz}
S.~Baek and P.~Ko, \emph{{Phenomenology of U(1)(L(mu)-L(tau)) charged dark matter at PAMELA and colliders}}, \href{https://doi.org/10.1088/1475-7516/2009/10/011}{\emph{JCAP} {\bfseries 10} (2009) 011} [\href{https://arxiv.org/abs/0811.1646}{{\ttfamily 0811.1646}}].

\bibitem{Heeck:2011wj}
J.~Heeck and W.~Rodejohann, \emph{{Gauged $L_\mu - L_\tau$ Symmetry at the Electroweak Scale}}, \href{https://doi.org/10.1103/PhysRevD.84.075007}{\emph{Phys. Rev. D} {\bfseries 84} (2011) 075007} [\href{https://arxiv.org/abs/1107.5238}{{\ttfamily 1107.5238}}].

\bibitem{Das:2013jca}
M.~Das and S.~Mohanty, \emph{{Leptophilic dark matter in gauged $L_{\mu}-L_{\tau}$ extension of MSSM}}, \href{https://doi.org/10.1103/PhysRevD.89.025004}{\emph{Phys. Rev. D} {\bfseries 89} (2014) 025004} [\href{https://arxiv.org/abs/1306.4505}{{\ttfamily 1306.4505}}].

\bibitem{Heeck:2014zfa}
J.~Heeck, \emph{{Unbroken B \textendash{} L symmetry}}, \href{https://doi.org/10.1016/j.physletb.2014.10.067}{\emph{Phys. Lett. B} {\bfseries 739} (2014) 256} [\href{https://arxiv.org/abs/1408.6845}{{\ttfamily 1408.6845}}].

\bibitem{Borgohain:2020csn}
H.~Borgohain and D.~Borah, \emph{{Survey of Texture Zeros with Light Dirac Neutrinos}}, \href{https://doi.org/10.1088/1361-6471/abde9b}{\emph{J. Phys. G} {\bfseries 48} (2021) 075005} [\href{https://arxiv.org/abs/2007.06249}{{\ttfamily 2007.06249}}].

\bibitem{Kallosh:1995hi}
R.~Kallosh, A.~D. Linde, D.~A. Linde and L.~Susskind, \emph{{Gravity and global symmetries}}, \href{https://doi.org/10.1103/PhysRevD.52.912}{\emph{Phys. Rev. D} {\bfseries 52} (1995) 912} [\href{https://arxiv.org/abs/hep-th/9502069}{{\ttfamily hep-th/9502069}}].

\bibitem{Anamiati:2019maf}
G.~Anamiati, V.~De~Romeri, M.~Hirsch, C.~A. Ternes and M.~T\'ortola, \emph{{Quasi-Dirac neutrino oscillations at DUNE and JUNO}}, \href{https://doi.org/10.1103/PhysRevD.100.035032}{\emph{Phys. Rev. D} {\bfseries 100} (2019) 035032} [\href{https://arxiv.org/abs/1907.00980}{{\ttfamily 1907.00980}}].

\bibitem{Giunti:1992hk}
C.~Giunti, C.~W. Kim and U.~W. Lee, \emph{{Oscillations of pseudoDirac neutrinos and the solar neutrino problem}}, \href{https://doi.org/10.1103/PhysRevD.46.3034}{\emph{Phys. Rev. D} {\bfseries 46} (1992) 3034} [\href{https://arxiv.org/abs/hep-ph/9205214}{{\ttfamily hep-ph/9205214}}].

\bibitem{DeGouvea:2020ang}
A.~De~Gouv\^ea, I.~Martinez-Soler, Y.~F. Perez-Gonzalez and M.~Sen, \emph{{Fundamental physics with the diffuse supernova background neutrinos}}, \href{https://doi.org/10.1103/PhysRevD.102.123012}{\emph{Phys. Rev. D} {\bfseries 102} (2020) 123012} [\href{https://arxiv.org/abs/2007.13748}{{\ttfamily 2007.13748}}].

\bibitem{Beacom:2003eu}
J.~F. Beacom, N.~F. Bell, D.~Hooper, J.~G. Learned, S.~Pakvasa and T.~J. Weiler, \emph{{PseudoDirac neutrinos: A Challenge for neutrino telescopes}}, \href{https://doi.org/10.1103/PhysRevLett.92.011101}{\emph{Phys. Rev. Lett.} {\bfseries 92} (2004) 011101} [\href{https://arxiv.org/abs/hep-ph/0307151}{{\ttfamily hep-ph/0307151}}].

\bibitem{Dev:2024yrg}
P.~S.~B. Dev, P.~A.~N. Machado and I.~Martinez-Soler, \emph{{Pseudo-Dirac Neutrinos and Relic Neutrino Matter Effect on the High-energy Neutrino Flavor Composition}},  \href{https://arxiv.org/abs/2406.18507}{{\ttfamily 2406.18507}}.

\bibitem{Esmaili:2009fk}
A.~Esmaili, \emph{{Pseudo-Dirac Neutrino Scenario: Cosmic Neutrinos at Neutrino Telescopes}}, \href{https://doi.org/10.1103/PhysRevD.81.013006}{\emph{Phys. Rev. D} {\bfseries 81} (2010) 013006} [\href{https://arxiv.org/abs/0909.5410}{{\ttfamily 0909.5410}}].

\bibitem{Bell:2019pyc}
N.~F. Bell, G.~Busoni and S.~Robles, \emph{{Capture of Leptophilic Dark Matter in Neutron Stars}}, \href{https://doi.org/10.1088/1475-7516/2019/06/054}{\emph{JCAP} {\bfseries 06} (2019) 054} [\href{https://arxiv.org/abs/1904.09803}{{\ttfamily 1904.09803}}].

\bibitem{Bell:2021fye}
N.~F. Bell, G.~Busoni, M.~E. Ramirez-Quezada, S.~Robles and M.~Virgato, \emph{{Improved treatment of dark matter capture in white dwarfs}}, \href{https://doi.org/10.1088/1475-7516/2021/10/083}{\emph{JCAP} {\bfseries 10} (2021) 083} [\href{https://arxiv.org/abs/2104.14367}{{\ttfamily 2104.14367}}].

\end{thebibliography}\endgroup
\end{document}